\documentstyle[12pt,aaspp4]{article}

\newcommand{\citeauthoryear}[3]{#1 #3}
\newcommand{\citeN}[1]{\renewcommand{\citeauthoryear}[3]{##1 (##3)}\cite{#1}}
\newcommand{\citeNP}[1]{\renewcommand{\citeauthoryear}[3]{##1 ##3}\cite{#1}}

\newcommand{\text}[1]{{\rm #1}}
\newcommand{\comment}[1]{}

\begin{document}

\title{Hydrodynamic and N-body Schemes On An Unstructured, Adaptive Mesh
with Applications to Cosmological Simulations}

\author{Guohong Xu}
   \affil{
   Board of Studies in Astronomy and Astrophysics, \\
   University of California at Santa Cruz, \\
   Santa Cruz, California 95064 \\
   email: xu@ucolick.org
        \  \\
        {\it Submitted at} \today, {\it Revised at}..........
}

\begin{abstract}
The theory and application of numerical methods for unstructured meshes
have been improved significantly in recent years.
Because the grids can be place arbitrarily in space, unstructured
meshes can provide much higher spatial resolution than regular meshes.
The built-in nature of mesh adaptivity for unstructured meshes gives
one way to simulate highly dynamic, hierarchical problems involving both
collisionless dark matter and collisional gas dynamics.
In this paper, we describe algorithms to
construct unstructured meshes from a set of points with periodic
boundary conditions through Delaunay triangulation,
and algorithms to solve hydrodynamic and N-body problems on an unstructured mesh.
A combination of a local transformation algorithm and the traditional
Bowyer-Watson algorithm gives an efficient approach to perform Delaunay
triangulation.
\comment{
We also introduce a new error estimator to avoid floating point round-off
error in geometric computation. }
A novel algorithm to solve N-body equations of motion on an unstructured mesh
is described.
Poisson's equation is solved using the conjugate gradient method.
A gas-kinetic scheme based on the BGK model to solve Euler equations is
used to evolve the hydrodynamic equations.
We apply these algorithms to solve cosmological settings,
which involve both dark and baryonic matter.
Various cooling and heating processes for primordial baryonic matter are
included in the code.
The numerical results show that the N-body and hydrodynamic algorithms based on
unstructured meshes with mesh refinement are well-suited for
hierarchical structure formation problems.

{\it subject headings} numerical methods, cosmology, galaxy formation.
\end{abstract}

\section{Introduction}

Numerical simulations in astrophysics turn out to be very challenging
because of the large dynamical range required in three dimensions.
Examples include modeling of star forming regions and the origin of galaxies.
In cosmology, structures are believed to have formed hierarchically,
requiring a simultaneous modeling of
structures on scales of $\sim 100$ Mpc and $\sim 10$ kpc.
Various hydrodynamical techniques have been explored to achieve such a
large dynamic range,
from Eulerian methods using regular meshes
(c.f. \citeNP{Cen92} and \citeNP{ROKC93})
and recently with  
Adaptive Mesh Refinement (AMR, c.f. \citeNP{BC89}, \citeNP{KCM92})
to Lagrangian methods like Smoothed Particle Hydrodynamics 
(SPH, c.f. \citeNP{HK89}, \citeNP{KWH96}).
Eulerian schemes without AMR are inadequate because of computational
expense and are wasteful because high resolution is typically not required
at all points in a simulation volume.
SPH methods provide higher spatial resolution than
regular mesh Eulerian methods such as TVD and PPM,
but have poorer shock resolution than shock capturing methods.

N-body codes can be classified as direct, in the case of
Particle-Particle methods and TREE methods
(\citeNP{BH86}, \citeNP{Hernquist87}),
or grid-based, such as the Particle-Mesh method (\citeNP{EDFW85}),
or hybrids, such as P$^3$M (\citeNP{HE81}) or TPM (\citeNP{Xu95a}),
depending on the potential solver.
In cosmological simulations involving gas and dark matter
it is desirable that the N-body and hydro solvers achieve similar spatial
resolution.
Normally, Eulerian schemes for N-body and gas are combined, such as PM+TVD
code (\citeNP{ROKC93}), and Lagrangian schemes go together, such as TREESPH
(\citeNP{HK89}).
Numerically it is appreciate to employ similar algorithms
for N-body and gas dynamics.

Recently, unstructured meshes have become increasingly popular in many
fields, such as geophysics, seismology, structural
mechanics and computational fluid dynamics.
When combined with an accurate shock-capturing technique, codes employing
unstructured meshes have many advantages over particle based algorithms and
Eulerian codes with regular grids, in principle.
In an unstructured mesh, grid points are connected
by triangles in two dimensions and tetrahedra in three dimensions.
Since grid points can be placed arbitrarily, an optimal mesh can
be configured for any applications.
Mesh refinement can be achieved by simply
adding more grid points and reconnecting the mesh.
The refined mesh will have the same topology as the original one except that
it will have more cells, thus mesh refinement adds no overhead
to algorithms designed for unstructured meshes.
Although the nodes in an unstructured mesh can be irregularly
distributed, the internal data structure used to represent the grid is
homogeneous as opposed to block-structured grids where 
block boundaries and interiors must be distinguished.
As described below, only local grid operations are needed to solve
equations on unstructured meshes, hence the codes can be easily
parallelized.

In this paper, we describe techniques to construct unstructured
meshes and algorithms to solve N-body and gas dynamic problems on these
grids.
We apply this code to perform cosmological N-body +
hydrodynamic simulations, including
various cooling and heating processes and mesh refinement.
To avoid lengthy mathematical derivations of the algorithms, we put 
all the required formulas in the appendices.

\section{Numerical Algorithms on Unstructured Meshes}

\subsection{Construction of Unstructured Meshes}

\begin{figure}[b]	
\epsfbox{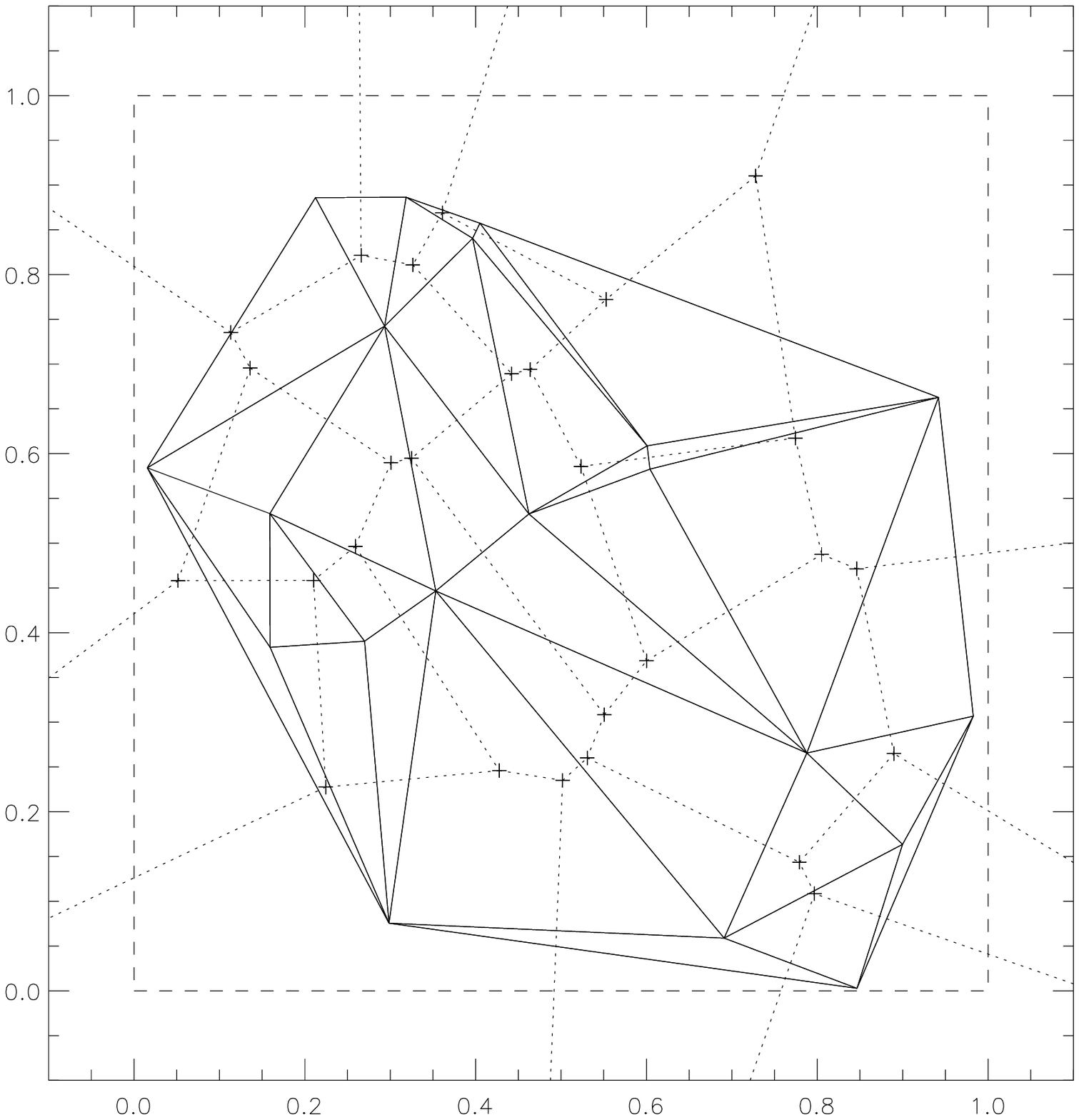}
\caption{\label{fig:voronoi}
Relation between Dirichlet tessellation (dashed lines)
and Delaunay triangulation (solid lines) in 2-dimensional space.
The crosses indicate the center of circumcenters of the corresponding
triangles.
}
\end{figure}

Unstructured meshes are constructed from arbitrarily scattered points in
$n$-dimensional space by Delaunay triangulation.
The mathematical definition of Delaunay triangulation can be found in text
books on geometric design.
An example of Delaunay triangulation in 2-dimensional space is illustrated
in Figure \ref{fig:voronoi}.
The scattered points are connected by non-overlapping triangles
obeying certain rules.
The interior of the circumcircle of any triangle 
contains no other point in the point set.
Delaunay triangulation is unique provided that no $n+2$ points
are co-spherical in $n$-dimensional space.
There are many properties associated with Delaunay triangulation
(c.f. \citeNP{Lawson86}, \citeNP{Barth95}), many of which will be cited in
our paper without strict mathematical description.
In Appendix A, we gave the essential formulas for geometric relations
between a point and a simplex (triangle in 2D and tetrahedron in 3D).

Much effort has been devoted to designing
algorithms to perform Delaunay triangulation.
Among these, incremental insertion algorithms are of particular interest,
since we will incorporate mesh refinement to enhance resolution.
The methods of Bowyer \cite{Bowyer} and Watson \cite{Watson}
are very similar.
When a new point is inserted into the existing triangulation,
those simplices with their circumspheres enclosing
the new point are deleted and and new simplices corresponding to those just
deleted are added.
The Bowyer-Watson algorithm is straight-forward to implement and is efficient
($O(N^{1+1/n})$ in $n$-dimensional space).
Another triangulation method is the edge swapping algorithm of
Green and Sibson \cite{GS77}, which was extended to 3D by Joe \cite{Joe89}.
Edge swapping algorithms insert a new point into the simplex
that encloses it, and perform a sequence of local transformations
until no further local transformations can be performed.
The edge swapping algorithm is slightly slower (also $O(N^{1+1/n})$),
but can be adopted to handle different criteria to perform
local transformations.

\begin{figure}
\epsfbox{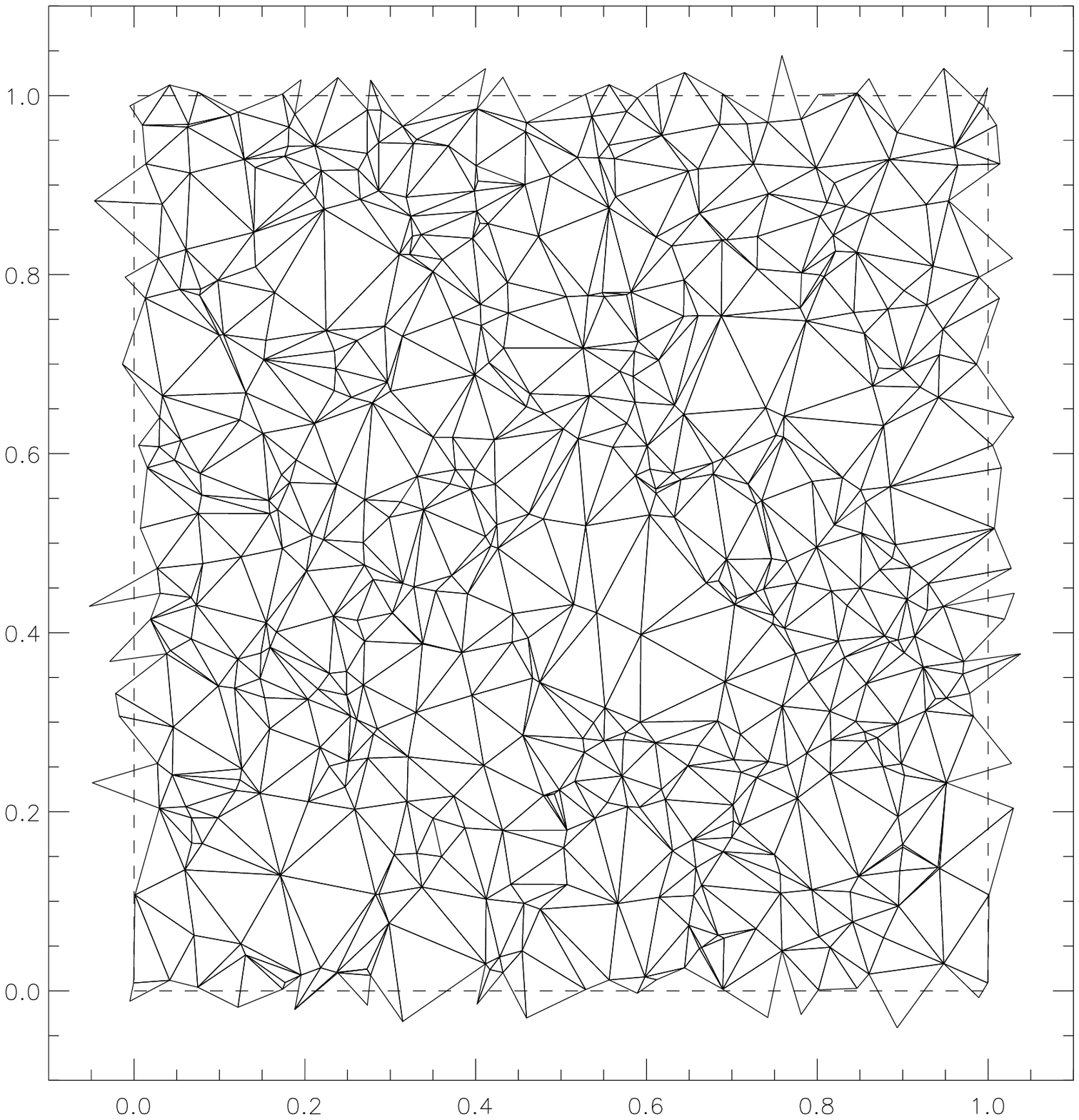}
\caption{\label{fig:umesh2d}
Example of 2-dimensional unstructured mesh constructed
by Delaunay triangulation. $1000$ nodes are randomly scattered in a
periodic box whose borders are indicated by dashed lines.
}
\end{figure}

For cosmological simulations, we require a triangulation algorithm
for a set of points in a periodic box.
The Bowyer-Watson algorithm appears to have difficulties with periodic
boundaries when the number of points in the box is small,
because some triangles might have two vertices that are periodic images of
the same point.
The edge swapping algorithm, however, can be applied to periodic volumes.
Thus a combined Bowyer-Watson algorithm with edge swapping
algorithm is efficient for periodic boxes.
In Figure \ref{fig:umesh2d} we show an example of a 2-dimensional
unstructured mesh for 1000 random points in a periodic box.

Our data structures to describe the geometric connections in an unstructured
mesh differ from those used previously (e.g. \citeNP{KV93}).
Two data types represent an unstructured mesh: {\it nodes} and {\it cells}.
A node contains a vector to describe the position of a point and a flag to
record the refinement level.
A cell contains $n+1$ pointers to its vertices, another $n+1$ pointer to its
neighbor simplices, and an integer flag to record various information about
the cell's status and its relation to its neighbors.
With periodic boundaries, another two integer flags record
the relative position between the simplex and its vertices and its neighbor
simplices.
So the memory requirement with periodic boundaries is
$n+1$ words for each node and $2n+5$ words for each cell.
This implies that the total memory required to store
an unstructured mesh is about
$3 + 2 \times 9 = 21$ words per node in 2-D, and
$4 + 6 \times 11 = 70$ words per node in 3-D.

\subsection{N-body algorithms on unstructured meshes}

Gravitational accelerations in N-body systems can be calculated
either from particle-particle methods or particle-mesh techniques.
Below we introduce a new particle-mesh algorithm for unstructured
grids. 

\begin{figure}
\epsfbox{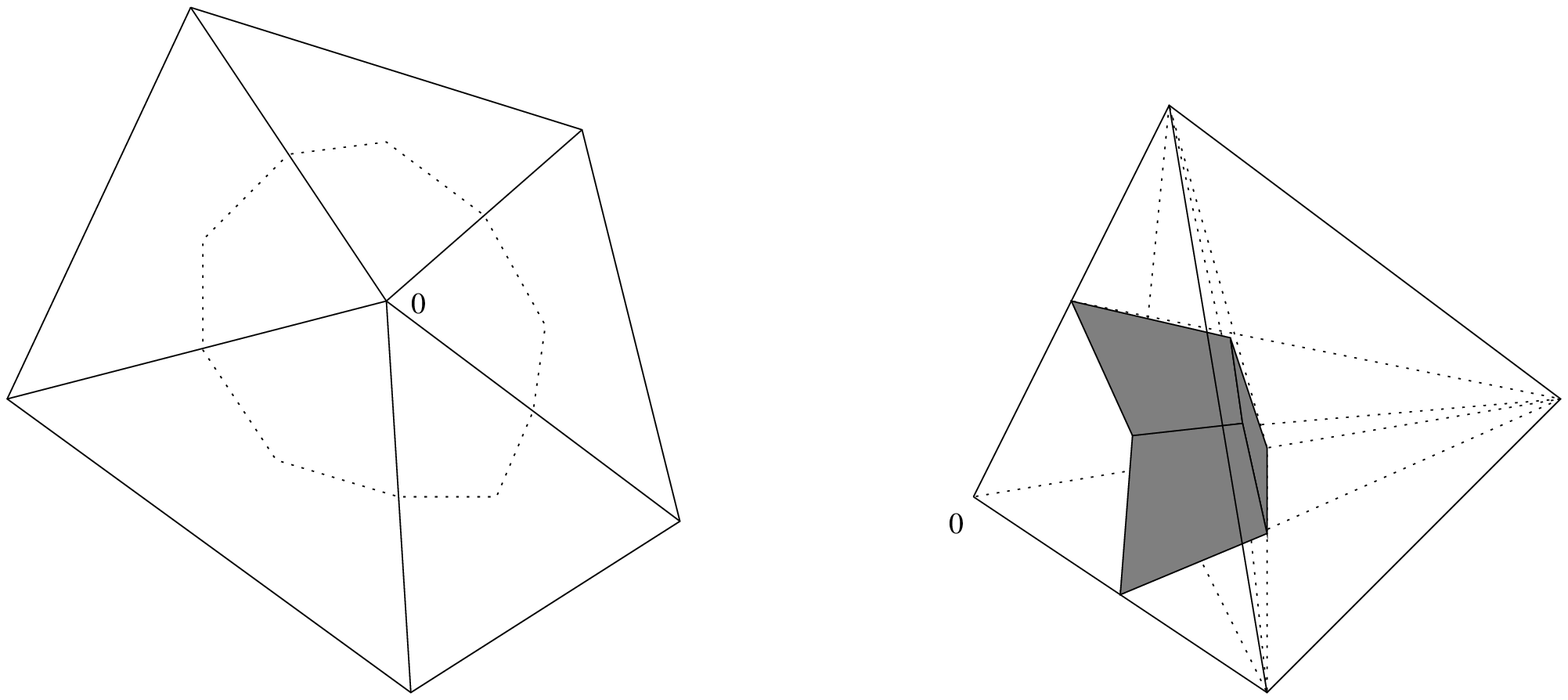}
\caption{\label{fig:dual}
Illustration of control volume of node 0 with 2-D case on the left and 3-D
case on the right.
On the left panel, the dashed lines, which connects the centers of edges
and the middle of triangles, indicates the control volume for node 0.
On the right panel, only one tetrahedron associated with node 0 is shown.
The shaded region represents one peice of the control volume boundaries.
}
\end{figure}

The discretized Poisson's equation on an unstructured mesh can be derived
as follows.
Consider node 0 as illustrated in Figure \ref{fig:dual}.
We choose the control volume for node 0 with the boundary
connected by the middle points of the edges and middle point of each
simplex associated with node 0,
such as that indicated by the dotted lines in Figure \ref{fig:dual}.
Integrating Poisson's equations over this control volume gives,
\begin{equation}
\int_\Omega dV \nabla^2 \phi = \int_\Omega dV 4 \pi G (\rho - \rho_b) 
= 4 \pi G ( m_0 - \rho_b V_0 ) ,
\label{eqn:poisson}
\end{equation}
where $V_0$ and $m_0$ are the control volume and mass of node 0, respectively.
If we linearly interpolate the potential field in each simplex $T_i$,
we have
$\phi(x) = \sum_k w_k(x) \phi_k$,
where $w_k(x)$ is the weighting to vertex $k$.
It can be shown that $w_k(x)$ is equal to the barycentric coordinates
of a point located at $x$ relative to vertex $k$ of simplex $T_i$
(see Appendix A for more details).
Using Gauss' theorem, the left hand side of equation~(\ref{eqn:poisson}) can
be written as,
\begin{equation}
\int_\Omega dV \nabla^2 \phi =
\int_{\partial \Omega} d \vec{S} \cdot \nabla \phi =
\sum_{T_i} \vec{S_i} \cdot \nabla \phi 
= - \sum_{T_i} \frac{V_i}{n} \nabla w_0 \cdot
\sum_{k=0}^{n} \nabla w_k(x) \phi_k
,
\end{equation}
where $V_i$ is the volume of simplex $T_i$ and $n$ is the dimension of space.
This result is the same as that derived using the Garlerkin
finite element approximation (\citeNP{Barth95}).

\begin{figure}
\epsfbox{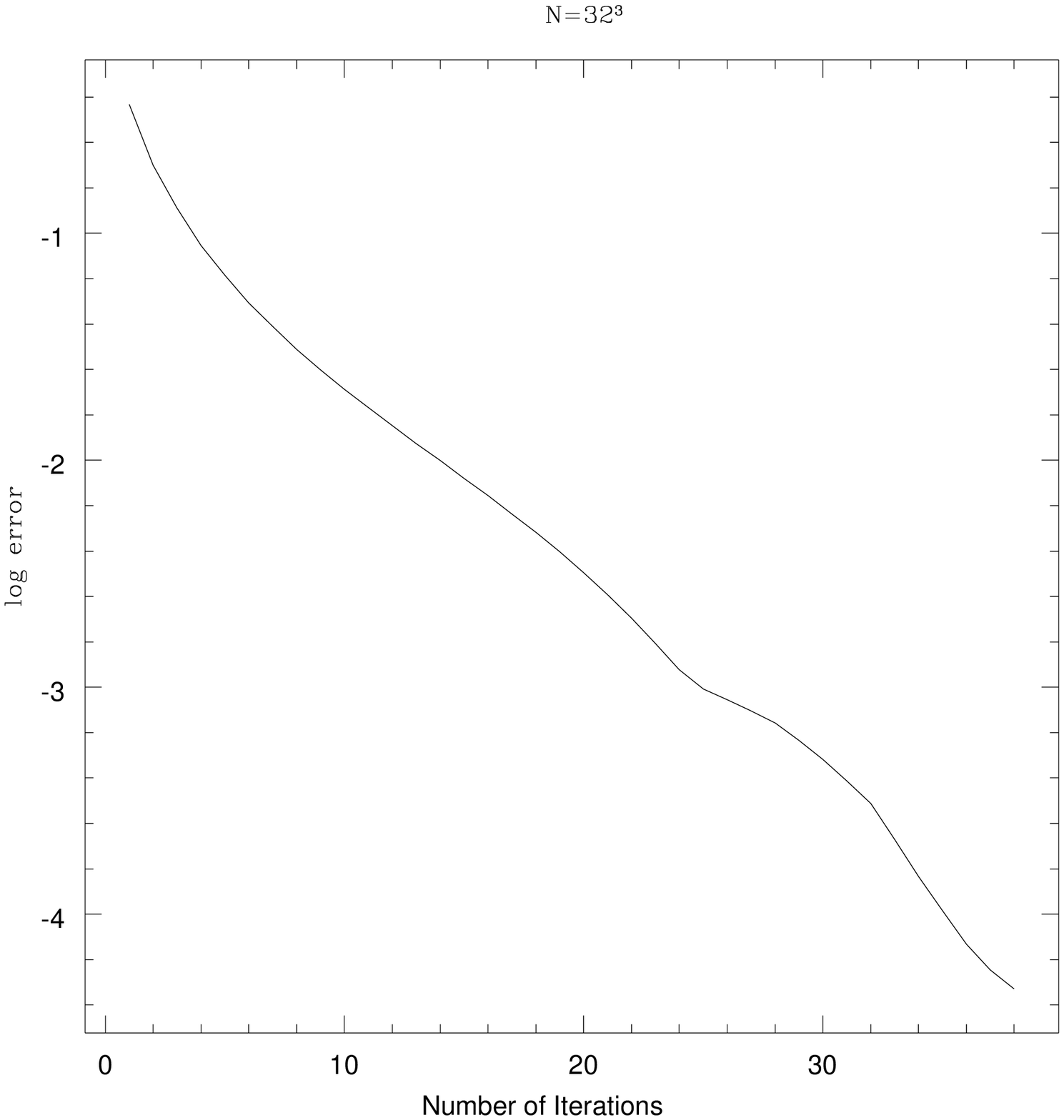}
\caption{\label{fig:cga}
The convergence rate of our conjugate gradient algorithm.
The error decreases almost exponentially as the number of iterasions.
}
\end{figure}

The discrete Poisson's equation resulting from this procedure is
a system of linear equations,
$M_{ij} \cdot \phi_j = b_i$, where $M_{ij}$ is a symmetric sparse matrix.
Since the matrix $M$ is sparse, the conjugate
gradient method can be used to solve this linear system efficiently.
To guarantee convergence, we require that $M$ be positive definite.
A necessary and sufficient condition
for $M$ to be positive definite is that the
triangulation is a Delaunay triangulation (\citeNP{Barth95}).
The convergence rate of this simple conjugate gradient method is
shown in Figure~\ref{fig:cga}.
The error decreases exponentially with the number of iterations,
given a good initial guess of the potential.
In N-body simulations, we can
always use the solution in the previous step as the initial guess.
Our numerical experiments indicate that the number of iterations required to achieve convergence during the next time step is typically about $20-50$. 

In order to solve Poisson's equation on an unstructured mesh for N-body
problem, we must interpolate particles to the mesh. For regular mesh, 
particle interpolation can be done using the Cloud-in-Cell (CIC)
interpolation (see, for example \citeNP{EDFW85}).
For an unstructured mesh, a similar procedure can be used.
For each particle, we identify the cell containing the particle,
and calculate the barycentric coordinate $b_i$ of the
particle inside the cell according to equation~(\ref{eqn:barycenter}).
Mass is assigned to each node of the tetrahedral cell with
weighing factor $b_i$.
After particle mass interpolation, the density of each node obtained
by dividing its mass by its dual volume.

After solving for the gravitational potential, the acceleration
on each node is calculated from the average of its control volume.
For example, the acceleration at node 0 in Figure \ref{fig:dual} is,
\begin{equation}
\vec{F_0} = \frac{1}{V_0} \int_\Omega (-\nabla\phi) dV
= \sum_{T_i} \frac{V_i}{V_0} \sum_{k=0}^{n} \phi_a \nabla w_k .
\end{equation}
The acceleration on each particle is calculated in a similar fashion as in the
Particle-Mesh algorithm,
\begin{equation}
\vec{F_i} = \sum_{k=0}^{n} w_k(\vec{x_i}) \vec{F_k} .
\end{equation}
The above formulation, when applied to a regular mesh, is identical to the
Particle-Mesh algorithm with CIC interpolation.

\begin{figure}
\epsfbox{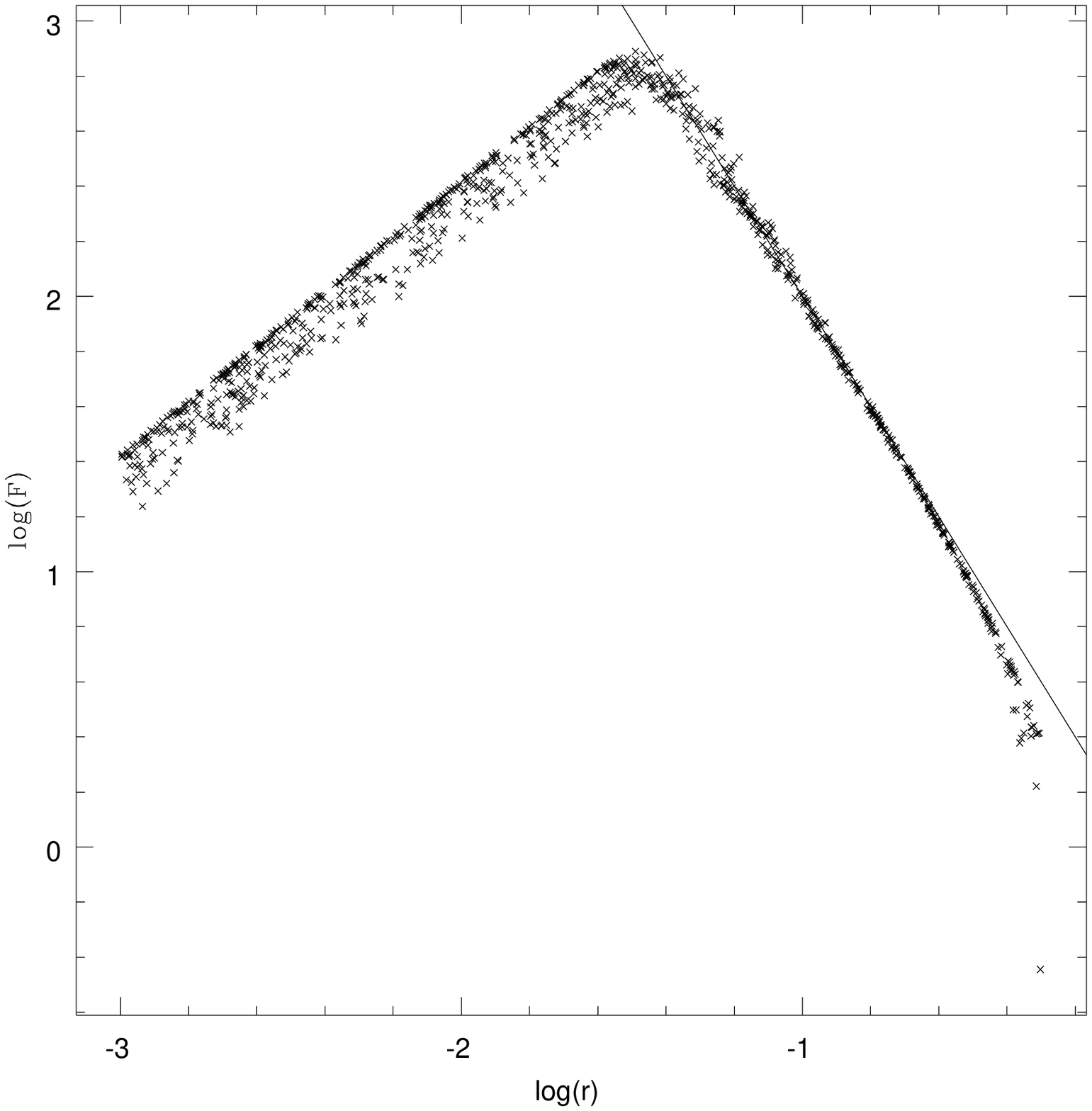}
\caption{\label{fig:force}
The force between two particles on an unstructured mesh 
using $32^3$ nodes.
The $32^3$ nodes distributed uniformly in a periodic box.
The solid line indicates the $r^{-2}$ law.
}
\end{figure}

In Figure \ref{fig:force}, we show the force between two equal mass particles
with different separations and orientation obtained using the above
algorithm.
The force behaves similarly to the PM method, i.e. accurate long distance
forces, but underestimate short range force.
The force is not very noisy within one cell, which
indicates the force resolution can be improved using techniques similar to
P$^3$M. The force resolution is roughly $1.5-2$ cells, which is slightly
better than the PM algorithm because each node is connected with 
more cells in an unstructured mesh than in a regular mesh.
The deviation from $r^{-2}$-law at large separations is due to periodic
boundaries.

As discussed later, mesh refinement can be performed at low cost,
so we can achieve high spatial resolution using unstructured
meshes for the N-body problem.

\subsection{Hydrodynamics on unstructured meshes}

The Euler equations are solved on unstructured meshes through the Finite Volume
scheme (c.f. \citeNP{VK94}),
\begin{equation}
\int_\Omega \frac{\partial U}{\partial t} d V
+ \int_{\partial \Omega} \vec{F} (U) \cdot d \vec{S} = 0 ,
\end{equation}
where $U \equiv \{ \rho, \rho \vec{v}, E \}^T$ represents the fluid state
and $\vec{F} (U)$ is the flux vector.
Before we write down the discrete form of the equations, we need to
decide whether we want to store the fluid variables on the nodes or in
the cells. The node representation uses less memory 
because the number of cells is typically $5-6$
times the number of nodes in 3-D. But there are indications that the cell
representation gives higher resolution than node representation
(see \citeNP{Mavriplis92} for a discussion).
We choose node representation to minimize memory usage.
Consider node 0 in Figure \ref{fig:dual} with its control volume
$\Omega$ illustrated in the figure. We have,
\begin{equation}
\left ( \frac{\partial U}{\partial t} \right )_0
= \frac{1}{V_0} \sum_{k \in {\mathcal N}_i} \vec{F}_{0k} \cdot \vec{S}_{0k} ,
\end{equation}
where ${\mathcal N}_i$ is the set of neighboring vertices connected with
node 0, $\vec{F}_{0k}$ is the flux at the middle of edge $0-k$, and
$\vec{S}_{0k}$ is the total surface area of the control volume related
to edge $0-k$. All we need is a method to calculate the flux in
the middle of each edge accurately.

Previous approaches to solve the Euler equations have used Upwind schemes
of various types to calculate the flux at the edges (c.f.
\citeNP{Barth95}).
Upwind schemes require artificial viscosity to achieve better
than first order accuracy.
Here we introduce a new approach based on the gas kinetic
theory of fluid dynamics.

The hydrodynamic equations (both the Euler equations and the
Navier-Stokes equations) can be derived from Boltzmann's equation
through the Chapman-Enskog procedure (see, for example, \citeNP{Shu91b},
Chapters 2 \& 3).
It is quite physical to derive numerical schemes for hydrodynamic
equations using the gas-kinetic theory.
Recently, \citeN{XP94} and \citeN{XMJ95} successfully developed
such numerical schemes based on the BGK model (\citeNP{BGK54}) of the
collisional Boltzmann's equation,
\begin{equation}
\frac{\partial f}{\partial t} + \vec{u} \cdot \nabla f =
\frac{g-f}{\tau} ,
\label{eqn:BGK0}
\end{equation}
where $g(t,\vec{x},\vec{u})$ is the equilibrium distribution function
and $\tau$ is the collisional time scale.
The BGK model accurately describes a large range
of situations, from very high density, high temperature flows
to very high Mach number ($>10^4$) flows.
The solution to this equation can be written as
\begin{equation}
f(t,\vec{x},\vec{u}) = {1 \over \tau} \int_0^t g(t',\vec{x}',\vec{u}) e^{-(t-t')/\tau} dt'
+ e^{-t/\tau} f_0 (\vec{x}-\vec{u} t, \vec{u})
\label{eqn:BGK1}
\end{equation}
where $\vec{x}' = \vec{x}-\vec{u} (t-t')$ and $f_0(\vec{x},\vec{u})$ is the initial state.

The macroscopic quantities $U(t,\vec{x})$ and $\vec{F} (t,\vec{x})$ are
moments of the distribution function $f(t,\vec{x},\vec{u})$,
\begin{eqnarray}
U_\alpha &=& \int \psi_\alpha f d \vec{u} d \xi \\
\vec{F}_\alpha &=& \int \vec{u} \psi_\alpha f d \vec{u} d \xi
\label{eqn:fluxdef}
\end{eqnarray}
where,
\begin{equation}
\psi_\alpha = \left ( \begin{array}{c} 1 \\ \vec{u} \\ {1 \over 2}
(\vec{u}^2 + \xi^2) \end{array} \right )
\end{equation}
with $\xi$ representing the internal variable with $K$ degrees of
freedom which will be discussed in Appendix B.

Consider an edge which connects two nodes. Without loss of generality,
we assume that the two ends of the edge have coordinates $(-1/2,0,0)$ and
$(1/2,0,0)$, and that the current time is $t=0$ and the time step is $\Delta T$.
the current position in consideration is $\vec{x}=0$.
Following the treatment in \citeN{XMJ95},
we can expand the distributions $f_0$ and $g$ along the $x$-direction
around the edge center as follows,
\begin{eqnarray}
f_0 (\vec{x},\vec{u}) &=& \left \{
\begin{array}{l}
g_0^L (1 + A^L x), \  x < 0 \\
g_0^R (1 + A^R x), \  x > 0
\end{array}
\right . \\
g (t,\vec{x},\vec{u}) &=& g_0^G \left \{
	\begin{array}{l} 1 + A^{GR} x + B t, \ x < 0 \\
		1+A^{GL} x + B t, \ x > 0
	\end{array} \right .
\end{eqnarray}
Here $g_0$ is the equilibrium state, which is a Boltzmann distribution
for hydrodynamic equations.
The coefficients of $A^L, A^R, A^{GL}, A^{GR}, B$ can be expanded
in velocity space as
$A^L = A^L_\beta \psi_\beta$, $A^R = A^R_\beta \psi_\beta$,
$A^{GL} = A^{GL}_\beta \psi_\beta$, $A^{GR} = A^{GR}_\beta \psi_\beta$,
$B = B_\beta \psi_\beta$,
with $A^L_\beta, A^R_\beta, A^{GL}_\beta, A^{GR}_\beta, B_\beta$
being constants.
The reason for the splitting of the right hand side and the left hand side
has its physical basis (\citeNP{XMJ95}) and serves as the mechanism
for shock capturing in the scheme.

Substituting the solutions to equation~(\ref{eqn:BGK1})
after the integration, we get,
\begin{eqnarray}
f(t,0,\vec{u}) &=& (1-e^{-t/\tau}) g_0^{L/R}
+ (-\tau + (t+\tau) e^{-t/\tau}) u_x A^G g_0^{L/R} \nonumber \\
	& & + ( t-\tau + \tau e^{-t/\tau} ) B g_0^{L/R} 
	+ e^{-t/\tau} f_0 (-\vec{u} t, \vec{u}) .
\label{eqn:BGK3}
\end{eqnarray}
This distribution is used to calculate the flux function
$\vec{F}_\alpha (t, \vec{x})$ through moment integration (equation
\ref{eqn:fluxdef}).

Notice in the above derivation that it does not matter if the gas is 1-D or
3-D.
\comment{
For the case of 3-D hydrodynamics, just make substitute $x$ to $\vec x$,
 $u$ to $\vec u$ and $V$ to $\vec V$. The whole formalism is still the same.
}
In this sense, the BGK gas-kinetic scheme is truly multiple dimensional 
without involving vector splitting which are usually used in TVD or PPM
codes to generalize from 1-D space to multiple dimensional space.

\section{Cosmological Equations}

The dynamical equations for dark matter and gas in a comoving frame 
can be written as follows (c.f. \citeNP{Peebles80}, \citeNP{Cen92}),
\begin{eqnarray}
\frac{d \vec{x_i}}{d t} &=& \frac{1}{a} \vec{v_i} \\
\frac{d \vec{v_i}}{dt} + \frac{\dot a}{a} \vec{v_i} &=& - \frac{1}{a} \nabla \phi \\
\nabla^2 \phi &=& \frac{4 \pi G}{a} (\rho - \rho_0) \\
\frac{\partial \rho}{\partial t} + \frac{1}{a}\frac{\partial \rho v_k}{\partial x_k} &=& 0 \\
\frac{\partial \rho v_j}{\partial t} + \frac{1}{a}\frac{\partial \rho v_j v_k}{\partial t} + \frac{1}{a} \frac{\partial p}{\partial x_j} &=& -\frac{\dot
a}{a} \rho v_j - \frac{1}{a} \rho \frac{\partial \phi}{\partial x_j} \\
\frac{\partial E}{\partial t} + \frac{1}{a} \frac{\partial (E+P)
v_k}{\partial x_k}
&=& -2 \frac{\dot a}{a} E - \frac{1}{a} \rho \vec{v} \cdot \nabla \phi
+ {\mathcal H} - \Lambda
\end{eqnarray}
The polytropic equation of state is normally adopted for an adiabatic gas,
\begin{equation}
E=\frac{1}{2} \rho \vec{v}^2 + \frac{p}{\gamma-1} .
\end{equation}
\comment{
The complicated terms in the above hydrodynamic equations make
it hard to apply the conventional hydrodynamic schemes to these equations
directly (see \citeNP{ROKC93} for a treatment).
We find the following transformation can restore the cosmological hydrodynamic
equations to their normal form.
By defining, 
}
These equations can be simplified by defining,
$ d t' =  a^{-2} dt$ ,
$ v_k' = a v_k $ ,
$ E' = a^2 E $ ,
$ p' = a^2 p $ and $\phi' = a^2 \phi$.
yielding,
\begin{eqnarray}
\frac{\partial \rho}{\partial t'} + \frac{\partial \rho v_k'}{\partial x_k} &=& 0
, \\
\frac{\partial \rho v_j'}{\partial t'} + \frac{\partial \rho v_j'
v_k'}{\partial x_k} + \frac{\partial p'}{\partial x_j} &=& - \rho \frac{\partial
\phi'}{\partial x_j} , \\
\frac{\partial E'}{\partial t'} + \frac{\partial (E'+p') v_k'}{\partial
x_k} &=& -\rho v_k' \frac{\partial \phi'}{\partial x_k}
+ a^4 ({\mathcal H} - \Lambda) , \\
E' &=& \frac{1}{2} \rho \vec{v}'^2 + \frac{p'}{\gamma-1} .
\end{eqnarray}

\subsection{Time Integration Scheme}

The equations of motion for the dark matter are integrated using
a time-centered, second order accurate leapfrog algorithm.
The particle positions are one half
time step ahead of their velocities.
\begin{equation}
\vec{x_i}^{n+1/2} = \vec{x_i}^{n-1/2} + a^{-1} \vec{v_i}^n \Delta t ,
\end{equation}
\begin{equation}
\vec{v_i}^{n+1} = \frac{1-\frac12 H(t) \Delta t}{1+\frac12 H(t) \Delta t} \vec{v_i}^{n}
+ a^{-1} {\vec F}_i^{n+1/2} \frac{\Delta t}{1 + \frac12 H(t) \Delta t} ,
\end{equation}
where ${\vec F} \equiv - \nabla \phi$.

In applying mesh refinement, we allow the system time step based on gravity
to adjust according to,
\begin{equation}
\Delta t_{grav} \le c_{grav} \min_{i,j}
\sqrt{ a^2 \delta l_{ij} \over \max(|{\vec F}_i|, |{\vec F}_j|)},
\end{equation}
where $\delta l_{ij}$ is the length of the edge between nodes $i$ and $j$
in the unstructured mesh, and ${\vec F}_i$ is the gravitational acceleration at
node $i$. The constant $c_{grav}$ has a meaning similar to the CFL condition
in hydrodynamics. Our numerical experiments show that
$c_{grav} \approx 0.3$ is a good choice.

When the time step changes from $\delta t_1$ to $\delta t_2$, we adjust the
particles positions from $t+\delta t_1/2$ to $t+\delta t_2/2$ using the
following second order accurate formula,
\begin{equation}
\vec{x}_i (t+\delta t_2/2) = \vec{x}_i (t+\delta t_1/2)
+ \frac{(\delta t_2-\delta t_1)}{2} \dot{\vec{x}_i} (t) 
+ \frac{(\delta t_2^2 - \delta t_1^2)}{8} \ddot{\vec{x}_i} (t-\delta t_1/2) .
\end{equation}

The Courant-Friedrichs-Lewy (CFL) stability criterion determines the
hydrodynamic time step for the system.
\comment{
For unstructured mesh, instead of
using the complicated CFL condition described in \citeN{Barth95} and
\citeN{VK94}, we use the following simplified version of CFL criterion,
}
We use simplified version of the CFL criterion in an unstructured mesh,
\begin{equation}
\delta t_\text{hydro} < \min_{ij} \left (
{ \delta l_{ij} \over
\max_{k=i,j} ( \vec{v}_k \cdot \vec{n}_{ij} + c_{s,k}) } \right ) ,
\end{equation}
where $\delta l_{ij}$ is the length of edge $i-j$, $n_{ij}$ is the
unit vector indicating the edge direction and $c_{s}$ is the sound speed.
We argue that the above criterion is sufficient to satisfy the CFL stability
criterion described in \citeN{Barth95}.

The gravitational terms in the cosmological hydrodynamic equations can be
solved consistently in the gas kinetic scheme by including the force term
in Boltzmann's equation. But it would be rather expensive to do so.
Instead, we treat these terms as source terms due to an external force.
The fluxes due to gravitational acceleration are calculated as follows,
\begin{eqnarray}
\Delta^G \rho &=& 0 , \\
\Delta^G \rho {\vec v}' &=& \frac12 (\rho^{(n+1)} + \rho^{(n)}) {\vec F}^{'(n+1/2)} \delta t' , \\
\Delta^G E' &=& \frac12 (\rho v^{'(n+1)} + \rho v^{'(n)}) {\vec F}^{'(n+1/2)} \delta t' ,
\end{eqnarray}
where ${\vec F}' \equiv -\nabla \phi'$.
Since the hydrodynamic quantities are synchronized with the velocity field
of the dark matter, when the system time step changes, we still need only
to change the particle positions.
For the hydrodynamic time step, we allow for variable CFL
constant from one time step to another in order to limit the change of system
time step. 

\subsection{Radiative Cooling}

\begin{figure}
\epsfbox{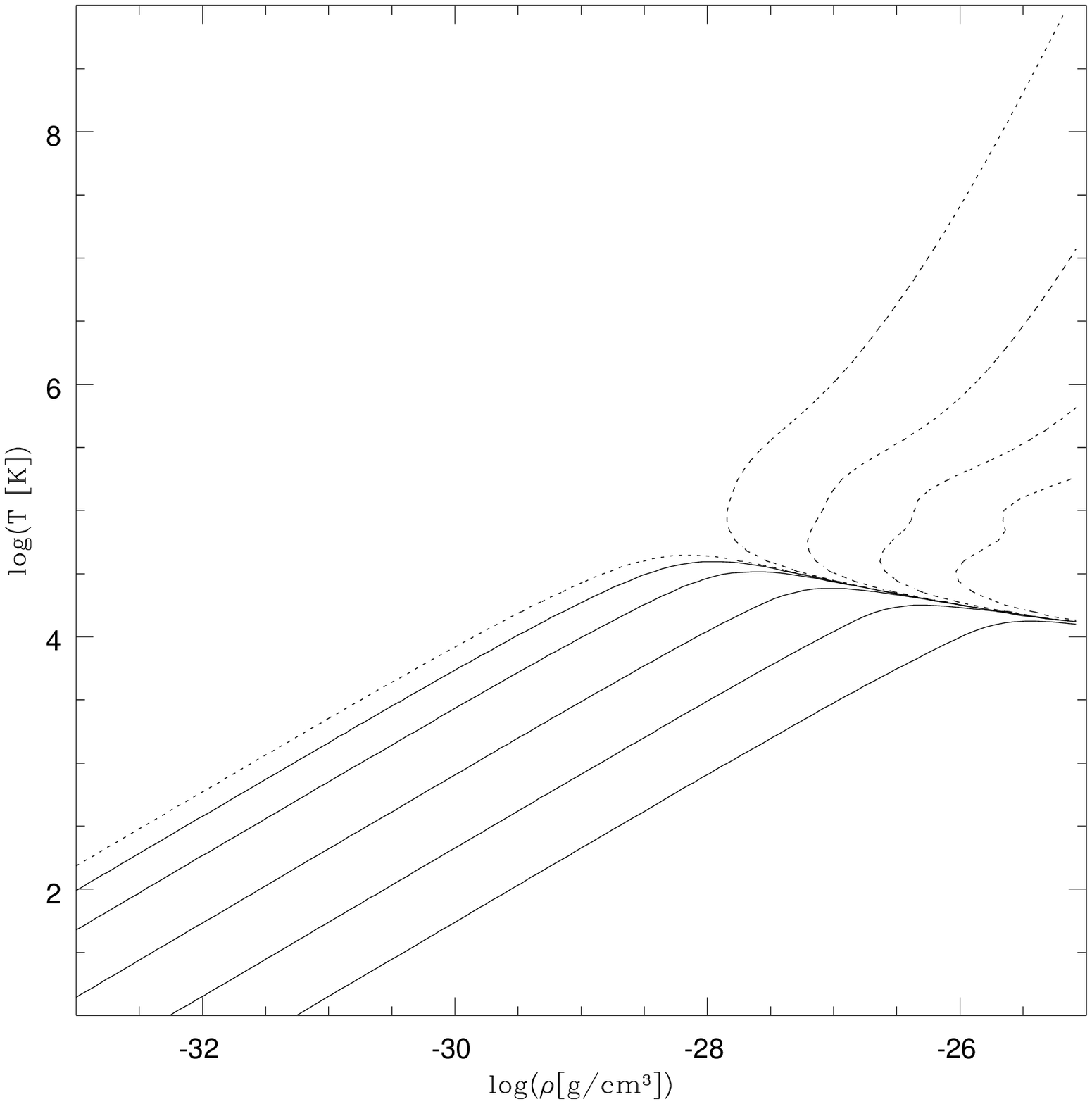}
\caption{\label{fig:cool_time}
Cooling time scale $t_\text{cool} = E/|\Lambda|$ with radiative cooling,
photoionization and inverse Compton cooling for gas at different density
and temperature. The UV radiation is assumed to be $J(v) = 10^{-22}
(\nu_L/\nu)$ erg/cm$^3$/sec/Hz, and the compton cooling is calculated at
$z=2$.
}
\end{figure}

Various cooling and heating processes relevant to primordial gas
have been included in the code (see Appendix D for a list of processes).
Since the cooling time can be very short compare with hydrodynamic time
(c.f. Figure \ref{fig:cool_time}), we have to be very careful with time integration of the energy
equation when cooling processes are turned on.
In our implementation, we integrated the cooling function with adjustable
time steps within one system time step. The variable step, fifth order
accurate Runge-Kutta integrator described in \citeN{NR92} is used to
integrate the following equation,
\begin{equation}
\frac{du}{dt} = \frac{\Delta u}{\Delta t} + {\mathcal{H}} - \Lambda  ,
\end{equation}
where $u$ is the thermal energy,
$\Delta u$ is the thermal energy change due to gravity and hydrodynamics, 
$\Delta t$ is the system time step
and $\Lambda$ is the cooling function. This equation is integrated from
0 to $\Delta t$ using many time steps depending on the cooling time
scale. Our numerical experiments show that sometimes about $10^4$ 
time steps is required within one dynamic time step $\Delta t$.

\subsection{Mesh Refinement Algorithms}

For an unstructured mesh, cells can be refined arbitrarily.
\citeN{VK94} discuss strategies for cell division.
One can put a node in the center of an edge, in the middle of a face,
in the middle of a cell, or a combination of all of them.

In cosmological simulations, we want to resolve forming structures.
For N-body problems, we use a mass criteria $m_c$ to determine mesh refinement.
After we interpolate particle data to mesh nodes, each node carries a mass.
For each face of a cell in the unstructured mesh, we put a refining node
at the center of the face if the linearly interpolated mass at the center
is above $m_c$.

For hydrodynamic problems, following the criteria for galaxy formation
in (\citeNP{CO93}), we put a
refining node in the middle of a cell if the gas in this cell,
(a) is contracting, $\nabla \cdot {\bf v} < 0$,
and (b) has a mass greater than the Jean's mass, $m_B > m_J$.
When cooling processes are included we also require the cooling time
to be shorter than the dynamical time,
$t_\text{cool} < t_\text{dyn}$.

\section{Code Tests and Performance}

\comment{
Based on unstructured mesh, we have now successfully derived algorithms to
solve N-body problem, hydrodynamic problem and cosmological equations
which involve both N-body processes and hydrodynamics.
Here, we want to using problems in cosmological structure formation to
demonstrate the performance of the above algorithms.
}

\subsection{Testing 1D gas-kinetic scheme}

\comment{
There have been many tests presented by \citeN{XMJ95} demonstrating the
capability of resolving shocks using the above gas-kinetic scheme for Euler
equations. Here, we would like to present a few more examples to test
that our implementation.
}

\begin{figure}
\epsfbox{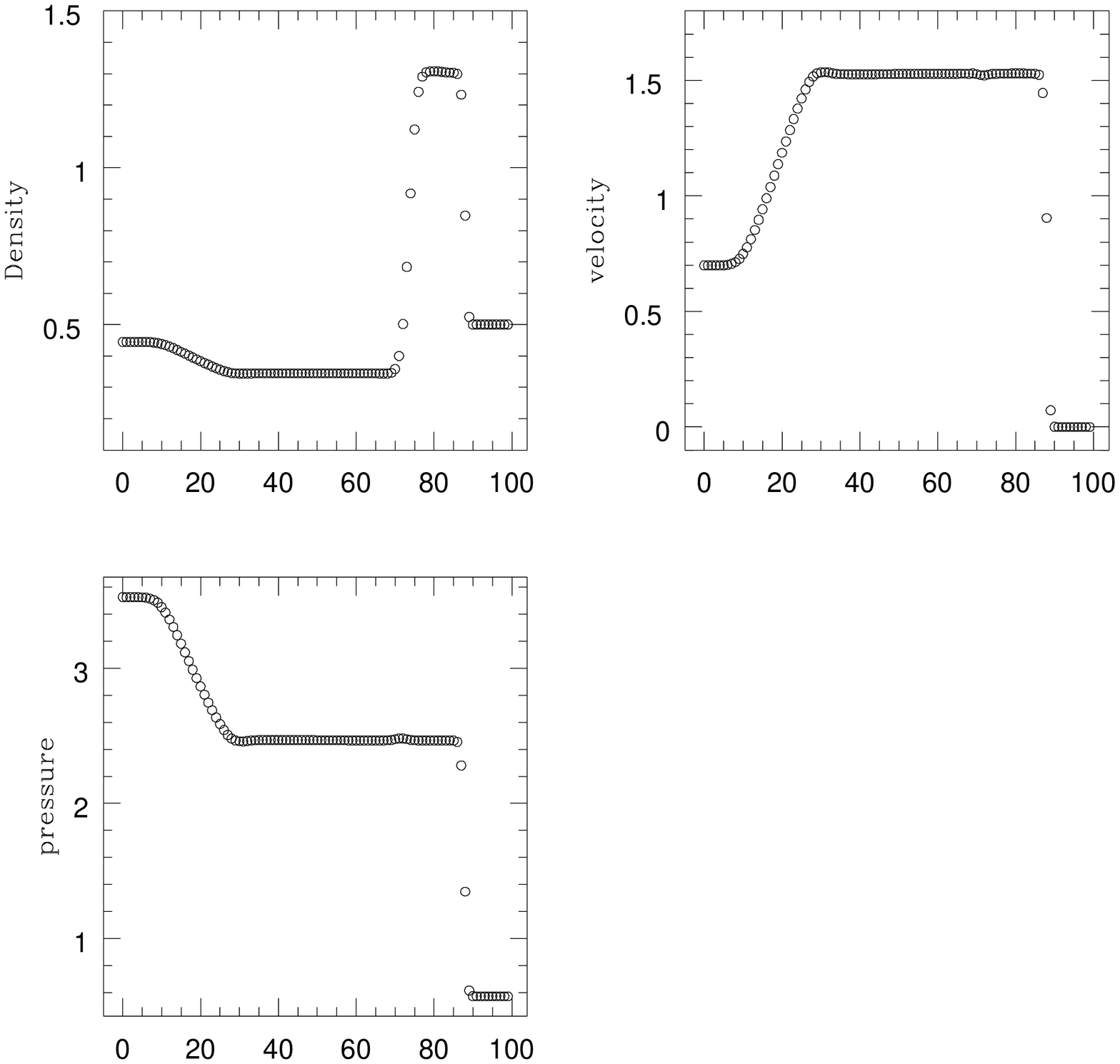}
\caption{\label{fig:shocktube2}
Lax shock tube test using the 1-dimensional BGK scheme. The result is at
time $t=0.15$. The BGK scheme captures shocks in two cells.
}
\end{figure}

In Figure \ref{fig:shocktube2}, we show the results of a Lax shock tube test.
The initial conditions for this test is $U=(0.445,0.311,8.928)$ for $x<0$ and
$U=(0.5,0,1.4275)$ for $x>0$.
The result is at $t=0.15$.
The contact discontinuity is resolved with about 2 cells,
the rarefaction shock was sharply captured with about two cells
and no post-shock oscillation is observed.

\begin{figure}
\epsfbox{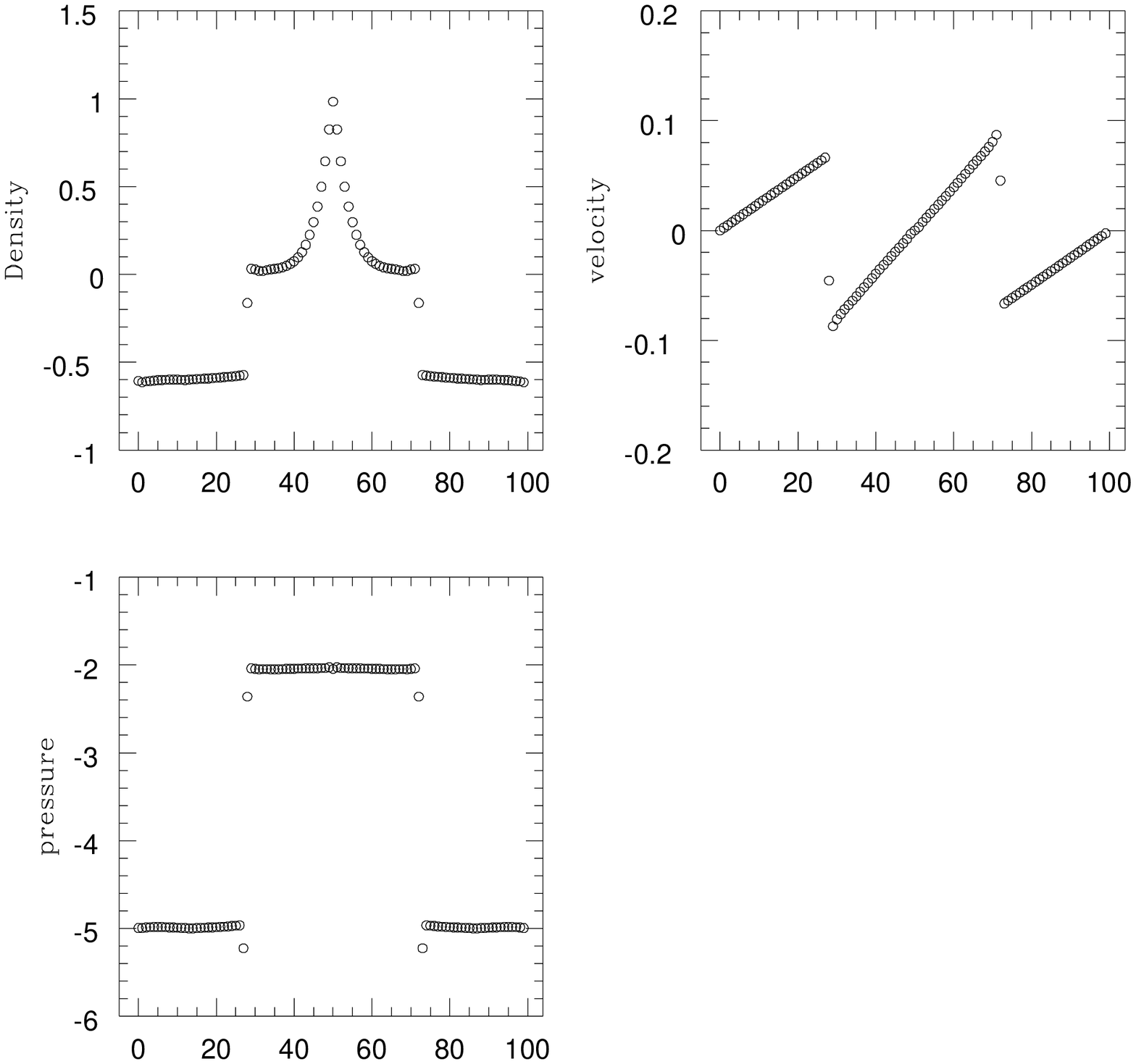}
\caption{\label{fig:caustic}
One dimensional caustic test with BGK gas-kinetic scheme using
$100$ zones. The results are at $t=3$.
}
\end{figure}

One test that is closely related to structure formation in cosmology is
described in \citeN{ROKC93} with the following initial conditions:
$\rho=0$, $v(x)=\sin(2 \pi x)/2\pi$, $p=10^{-4}$, and periodic boundary
condition for $0 \le x < 1$. Our results at $t=3$ are presented in Figure
\ref{fig:caustic}.
We notice that the BGK gas-kinetic scheme can successfully resolve features
within two cells without any artificial viscosity or adjustment
for the temperature term due to high Mach number.
Our scheme successfully reproduces the density
caustic, the saw-shape velocity field,
and segmented pressure field with small oscillation.
The Mach number involved in this test is much higher ($\sim 10^4$) than
that in usual shock tube tests.
This result demonstrates that the BGK gas-kinetic scheme is very
robust in high Mach number situations.

\subsection{Cosmological N-body Simulations}

\begin{figure}
\epsfbox{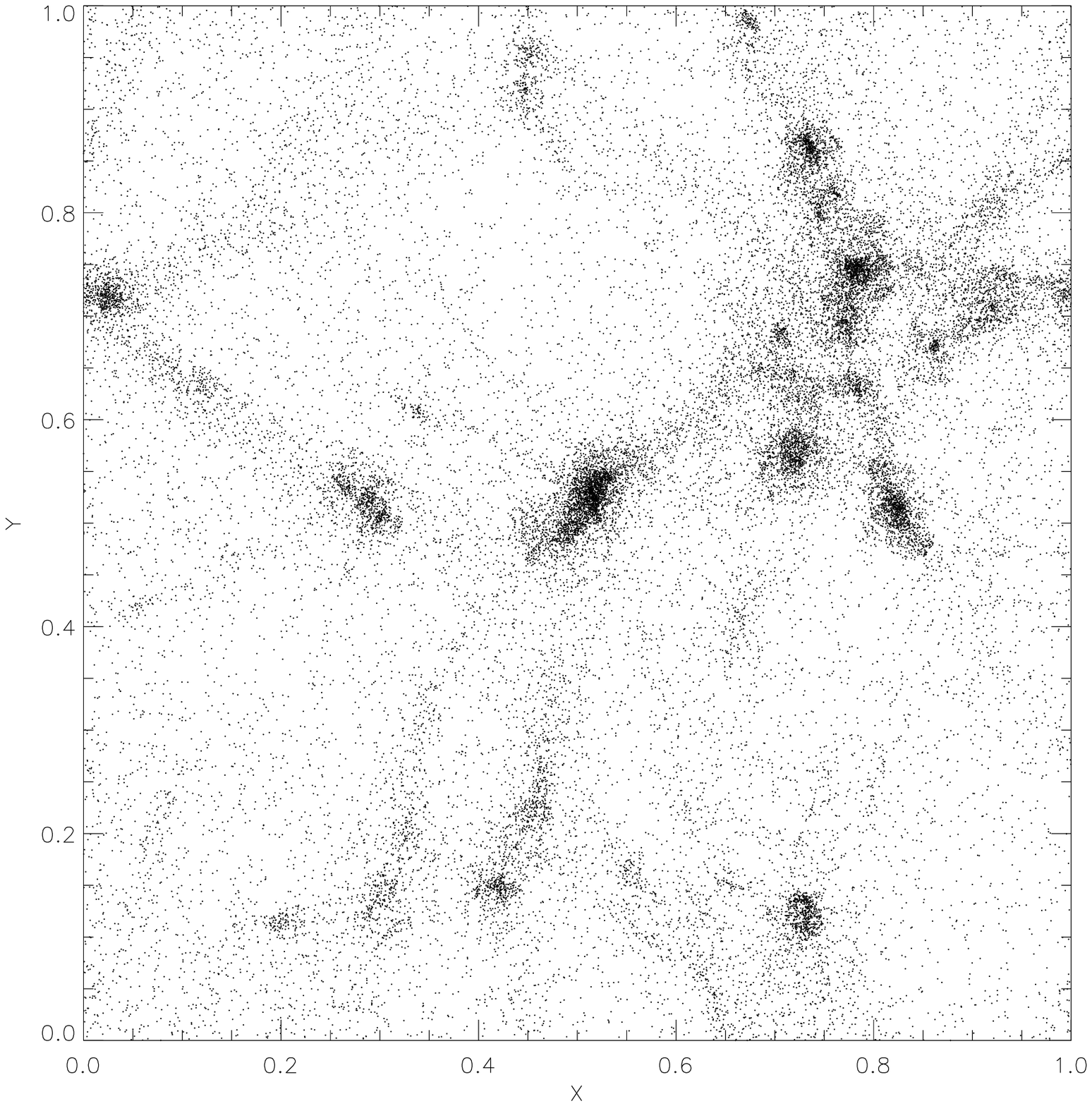}
\caption{\label{fig:dm_f32}
Result of a pure N-body unstructured mesh simulation without mesh refinement.
All the $32^3$ particles are projected in the X-Y plane. The raw mesh is
a $32^3$ uniform mesh.
}
\end{figure}

\begin{figure}
\epsfbox{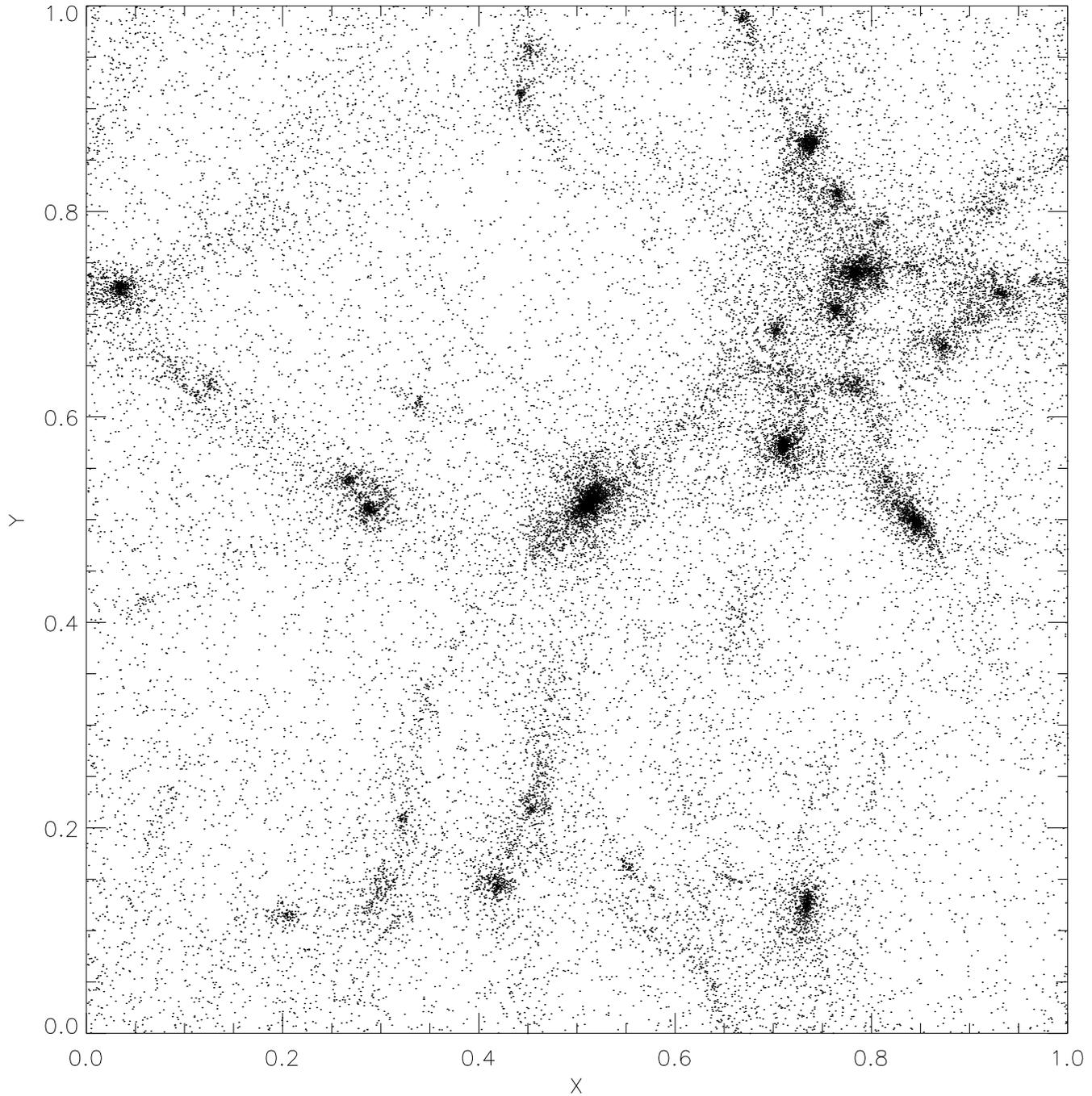}
\caption{\label{fig:dm_R5}
Result of a pure N-body unstructured mesh simulation with mesh refinement.
All the $32^3$ particles are projected in the X-Y plane. The raw mesh is
a $32^3$ uniform mesh. Refinement is done by a mass criterion $m_c = 5*m_i$,
where $m_i$ is the mass of each particle. 
}
\end{figure}

\begin{figure}
\epsfbox{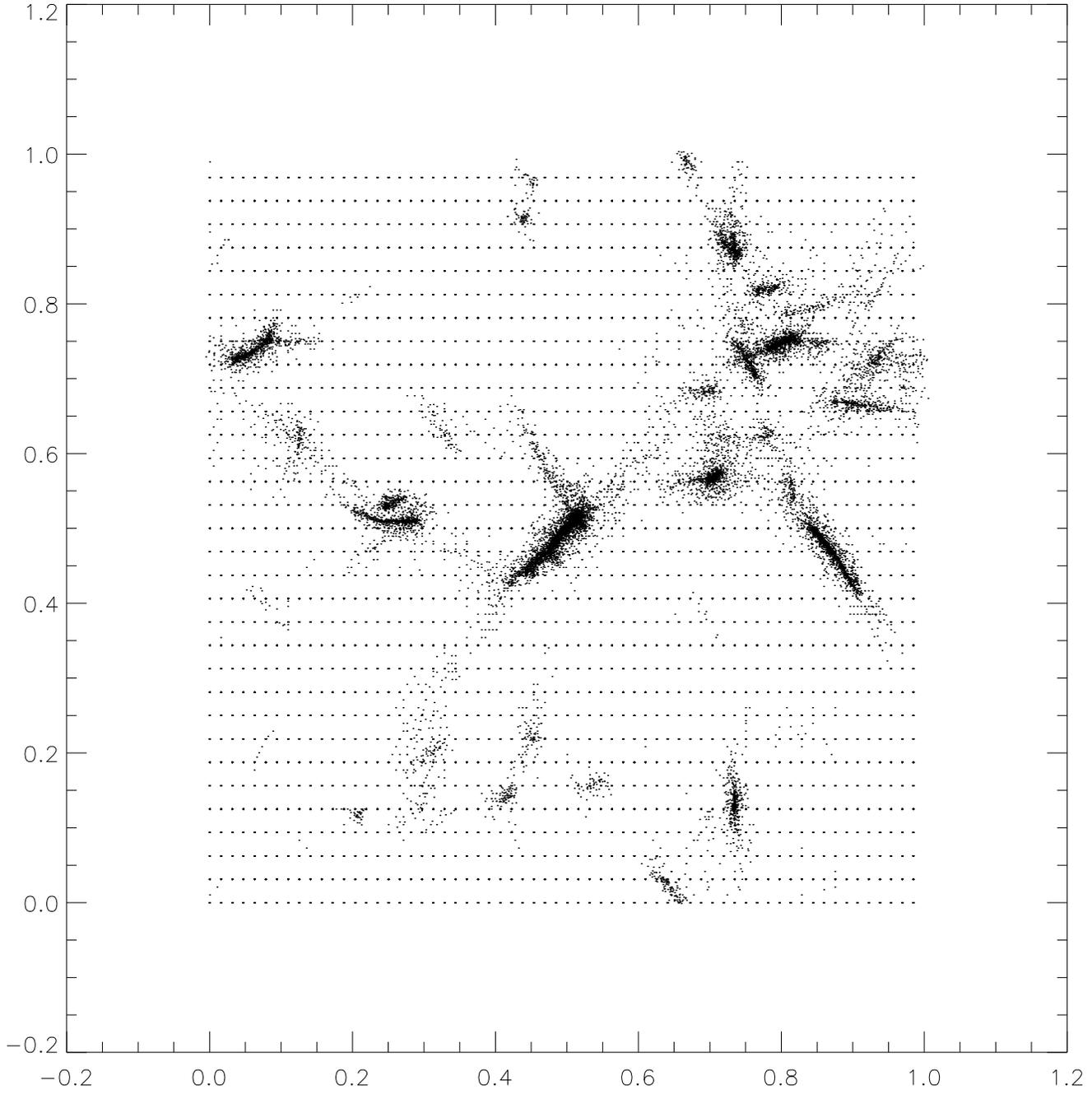}
\caption{\label{fig:um_R5}
Distribution of the mesh nodes projected to the X-Y plane.
The particle distribution of this simulation is shown in Figure \ref{fig:dm_R5}.
}
\end{figure}

\begin{figure}
\epsfbox{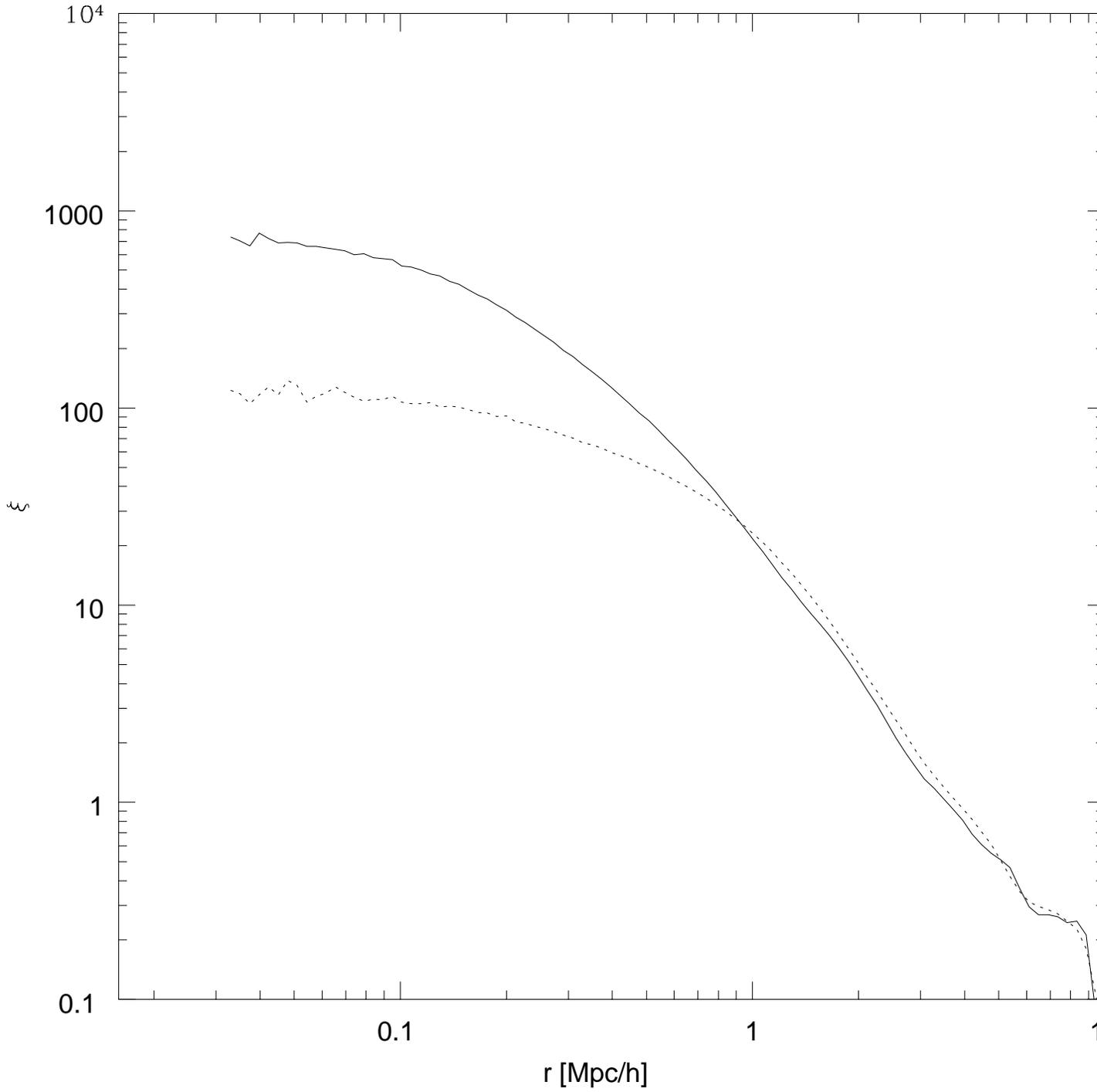}
\caption{\label{fig:corr}
The two-point correlation function $\xi(r)$ for two dark matter only simulations. The solid line is for the simulation with mesh refinement, and the dash line is without mesh refinement. In both simulations, $32^3$ particles are used on a raw mesh of
$32^3$ nodes.
The cell size for the raw mesh is $1$ Mpc/h.
}
\end{figure}

We construct the initial conditions for the Cold Dark Matter model using
the Zel'dovich approximation (cf. \citeNP{EDFW85}). Initially, particles are
almost uniformly distributed. As the system evolves, structure forms due to
gravitational clustering. More and more massive objects form as time passes
by. We show our results with a $32^3$ uniform mesh in Figure \ref{fig:dm_f32},
and results with mesh refinement from a $32^3$ uniform mesh in Figure
\ref{fig:dm_R5}. 
The final mesh nodes are shown in Figure \ref{fig:um_R5}, which indicates
that our mesh refinement traces the particle distribution very well.
Here, the mesh refinement is performed on the faces of tetrahedra. A new
node is put in the middle of a face if the mass on all the three nodes
are above a certain value.
In this test case, the critical mass is taken
to be $5 m_i$, where $m_i$ is the mass for each particle.
Visually, we can already see the great improvement of the resolution with
mesh refinement.
The two-body correlation function $\xi(r)$ (see \citeNP{Peebles80} for
a definition) for two simulations with and
without mesh refinement is shown in Figure \ref{fig:corr}.
The resolution improvement of the simulation with mesh refinement over the
simulation with mesh refinement is well above a factor of 10, while the
running time between the two simulations for each time step is less than a factor of two.

\subsection{Cosmological N-body + Gas Simulations}

\comment{
In an early study, \citeN{Comparison94} compared various cosmological
hydrodynamic codes available at the time, namely three Eulerian codes
and two SPH codes. The authors find dramatic differences between the
results from Eulerian codes and SPH codes.
}
Recently \citeN{Comparison96} have proposed a comparison 
between different cosmological hydrodynamic codes.
They set up a constrained initial conditions for a CDM model.
The initial conditions for the results described below are generated from
their density field using the Zel'dovich approximation (\citeNP{EDFW85}).
Readers can compare some of our results to others
presented in their paper (\citeNP{Comparison96}) with the same initial
conditions.
All the tests shown below are obtained with $32^3$ particles and $32^3$ raw
mesh with mesh refinement.

\begin{figure}
\epsfbox{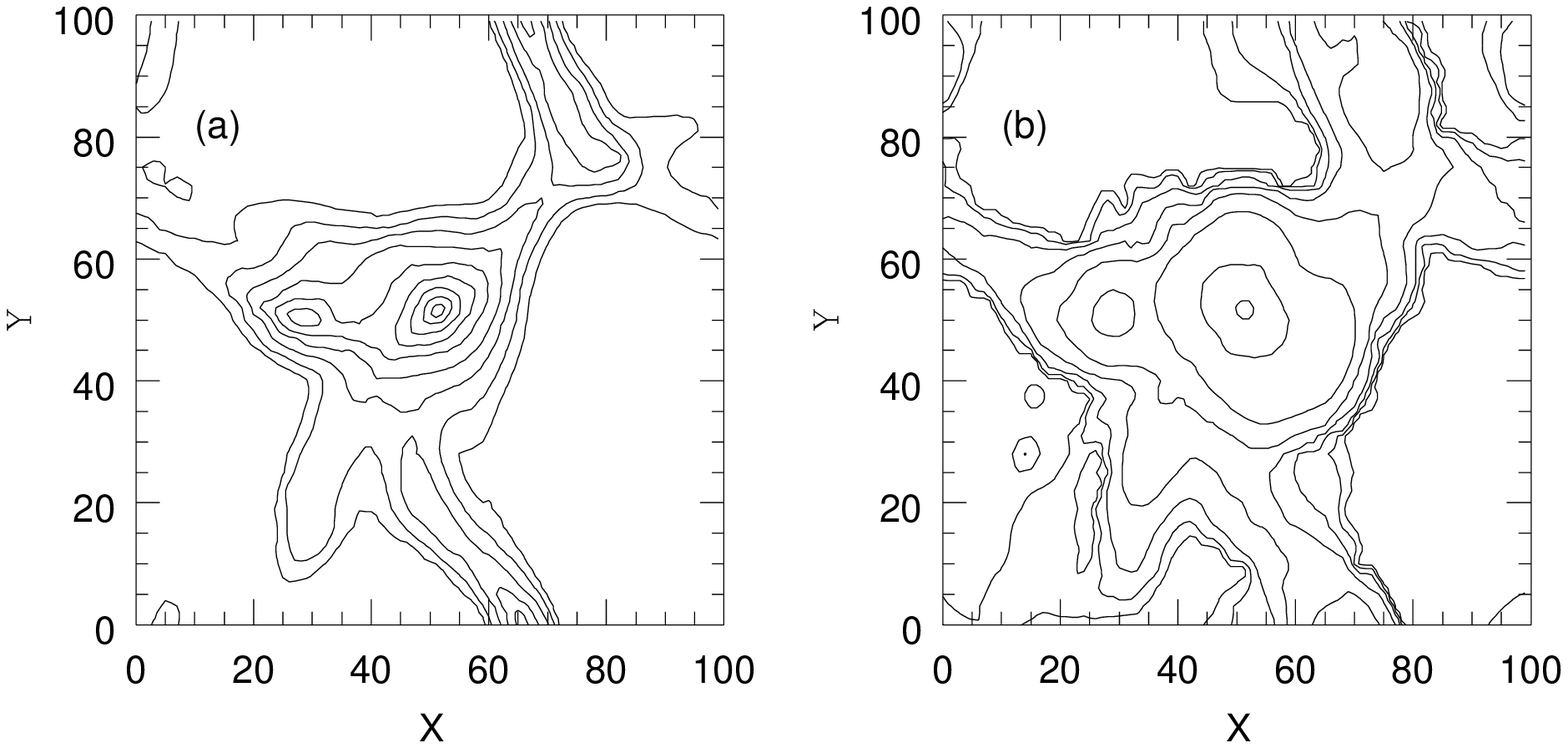}
\caption{\label{fig:slice}
Panel (a) shows the density contours of a slice in the simulation. The
contour levels are $\rho/<\rho> = 10^{k/4}$ with $k=0,1,\dots$.
Panel (b) shows the temperature contours of the same slice with contour
levels $T=10^{5+k/2} K$ with $k=0,1,2,\dots$.
}
\end{figure}

In Figure \ref{fig:slice}, we show the density and temperature contours
of a slice in the simulation. The density field shows some traces of
filamentary structure, and the temperature field is almost isothermal in
high density regions.

\begin{figure}
\epsfbox{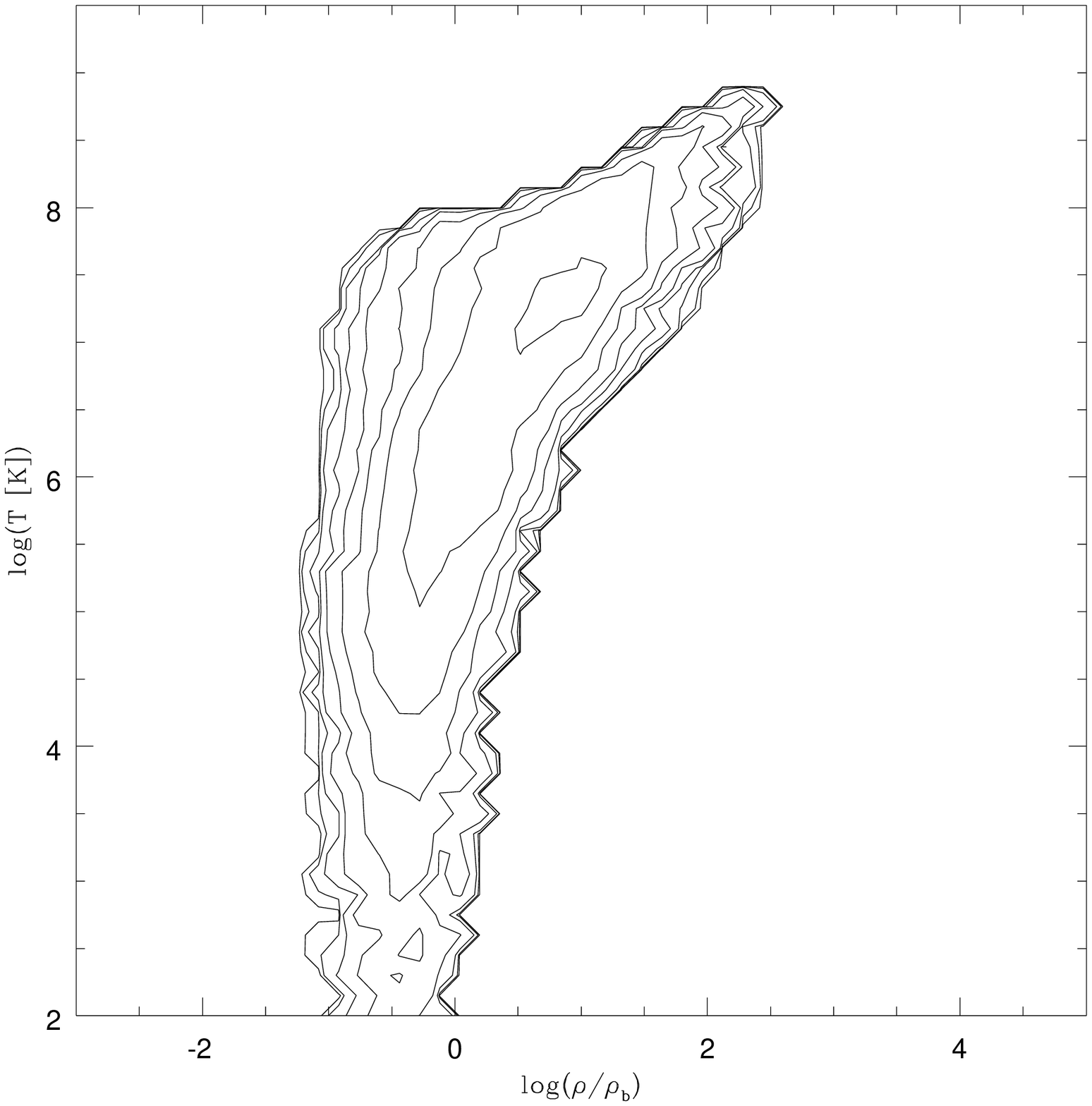}
\caption{\label{fig:rhoT}
Mass contour of gas at different density and temperature.
The total mass in the box is 1.
The contour levels are in $10^{i/2}$, $i=-4,-8$.
assumed to be 1
}
\end{figure}

In Figure \ref{fig:rhoT}, we show the fraction of mass contours of
baryonic matter with various density and temperature.
This figure summarizes the thermal state of the intergalactic medium.
From this figure we know that most of the baryonic matter stays at the average
density and in a temperature range of $10^4-10^6$ K. Only a small
fraction of the baryonic matter is in high density regions,
and that at high density regions remains at high temperatures ($\sim 10^7$ K).
The material at underdense regions is cold ($T \lesssim 10^5$ K).
The heating and cooling processes change the thermal state of intergalactic
medium dramatically.
One important feature is that
the gas at high density region can actually stay very cold. 
The line indicating the heating and cooling balance in Figure
\ref{fig:cool_time} indicates the thermal states of
these high density, low temperature gas.
Our simulations with heating and cooling processes actually have some
fraction of gas in these states.

\comment{
In Figure \ref{fig:profiles}, we show the dark matter and gas
density profiles of the central object in the box.
The dark matter profile is shown in solid line and the gas profile is shown
in dash line and shifted up by a factor of $10$ ($\Omega_B=0.1$).
The density ratio between dark matter and gas remains almost constant in
the outskirt of the cluster, but the dark matter is more concentrated in
the central region.

\begin{figure}
\epsfbox{figs/frenk_prof.ps}
\caption{\label{fig:profiles}
Density profiles of central object in a simulation with both dark matter
and baryonic matter.
The initial condition was generated by
\protect\cite{Comparison96}.
The Dark matter density profile is shown in solid line and the gas density
profile is shown in dashed line.
The gas density profile is moved up by a factor of $10$ in order to be
on the same scale as the dark matter profile.
As a comparison, the $\rho \propto r^{-2}$
power-law is shown in dotted line.
In the simulation, mesh refinement was performed using mass criterial on
both the dark matter and the gas respectively.
}
\end{figure}
}

\section{Summary and Discussion}

In this paper, we described a new cosmological N-body + gas dynamics code
based on algorithms for an adaptive, unstructured mesh.
The novel elements of this code are:
(1) the mesh construction;
(2) solving N-body systems;
(3) solving hydrodynamic equations;
(4) time step estimation and time integration;
(5) mesh refinement;
and 
(6) relevant heating and cooling processes for primordial gas.

The mesh construction with periodic boundary conditions
is performed using a combined Bowyer-Watson
algorithm and local transformation algorithm.
The initial mesh for cosmological simulations is a uniform, staggered mesh.
When some refining grids are required, new grid points are added to the mesh
structure through the incremental Bowyer-Watson algorithm, which modifies
the previous mesh structure slightly.
The incorporation of mesh refinement in unstructured mesh gives one way
to achieve high spatial resolution at relatively low cost.
For best results, a good refinement criterion is essential.
In general, refinement criterion can be derived both on physical
and numerical bases.

Poisson's equation is discretized on an unstructured mesh and solved using
conjugate gradient method.
Particles are interpolated to the mesh nodes using linear interpolation.
The resulting N-body algorithm is similar to the Particle-Mesh method with
CIC interpolation.
Because each node in an unstructured mesh has more associated cells, the
new N-body algorithm has slightly higher force resolution than the PM
algorithm.

We solve Euler's equations using finite-volume method.
Flux functions are calculated using BGK gas-kinetic scheme.
The gas-kinetic scheme constructs a time dependent distribution in the
middle of an edge and calculate flux functions by moments of the distribution
function.
This scheme provides high resolution for shock capturing and is very stable for high Mach number flows.

The new cosmological code solves dark matter and gas dynamics with the same
resolution. Our tests demonstrate that this code can provide high
spatial resolution by mesh refinement.
We include relevant cooling and heating processes for the primordial gas to
simulate the evolution of intergalactic medium accurately.

\comment{
In this paper, we described algorithms to perform N-body and hydrodynamic
simulations using an adaptive, unstructured mesh.
Based on strict mathematical derivation we have developed algorithms for
Delaunay triangulation with periodic boundary condition, and we have also
provide a new formula for sphere test which can estimate floating point
round-off errors.
We have developed methods to interpolate particles to the unstructured mesh,
and to solve Poisson's equation on unstructured mesh.
The gas-kinetic scheme based on BGK model for Euler equations is derived in
details. We applied the gas-kinetic scheme to the unstructured mesh and
build an algorithm to solve Euler equations using finite-volume scheme.

Our cosmological code with unstructured mesh can solve N-body and gas
dynamics with same resolution. The radiative processes for primordial
baryonic matter are also encoded.
This code can reach much higher resolution than the fixed grid Eulerian codes.
Together with the physical processes, this code can perform very
interesting cosmological simulations of structure formation.
}

\comment{
There is an interesting feature related to our N-body algorithm. In Figure
\ref{fig:force}, we show the force between two particles. Notice that the
force below $\sim 2$ cell size is much less noisy than the normal PM
algorithm using CIC interpolation, a correspond short range force function
can be derived to achieve sub-cell accuracy for pure N-body simulations
using unstructured mesh.
}

Since the internal data structures for an unstructured mesh are homogeneous,
unstructured mesh codes can be easily parallelized. The difficult part
of parallelization is the Delaunay triangulation procedure.
A parallel version of the code is being developed using the portable Message
Passing Interface (MPI) library functions.

The author would like to thank Michael Norman for suggesting the
investigation of unstructured mesh schemes, and Timothy J. Barth for kindly
providing many references and some discussions.
I greatly appreciate numerous detailed discussions with Kun Xu about the
gas-kinetic scheme and sharing opinions about computational fluid dynamics.
It is a great pleasure to thank Lars Hernquist and Jeremiah P. Ostriker for
their support and encouragement.
This work is supported by the NSF grant ASC 93-18185 rewarded to the Grand
Challenge Cosmology Consortium.

\appendix

\comment{
To define Delaunay triangulation strictly, we have to define the following
concept first. A {\it simplex} is the convex hull of $n+1$ points in
$n$-dimensional space, which are called {\it vertices} of the simplex.
A {\it triangulation}, $\mathcal{T}$ of a set of points $\mathcal{P}$ is a set
of non-degenerate simplices $\{T_i\}$ with the following properties:
\begin{enumerate}
\item All vertices of each simplex are members of $\mathcal{P}$.
\item The interiors of the simplices are pairwise disjoint.
\item Each facet of a simplex is either on the boundary of the convex hull
of $\mathcal{P}$, or else is the common facet of exactly two simplices.
\item Each simplex contains no points of $\mathcal{P}$ other than its
vertices.
\item The union of $\{T_i\}$ is exactly the convex hull of $\mathcal{P}$.
\end{enumerate}
The {\it Dirichlet tessellation} of a set of points $\mathcal{P}$ in
$n$-dimensional space are defined to be a set of cells $d_p$ which are
defined for each point $p \in \mathcal{P}$ such that all the points in cell
$d_p$ are closer to the point $p$ other than any other point of
$\mathcal{P}$.
The {\it Delaunay triangulation} of a set of points $\mathcal{P}$
is defined that is, in certain sense, be dual to the Dirichlet tessellation
of this set of points. To define how the Delaunay triangulation is dual to
Dirichlet tessellation requires more mathematics, which we want to skip
here. Instead, we show in Figure \ref{fig:voronoi} the Dirichlet
tessellation of 20 points in 2-dimensional space in dash lines, and the
Delaunay triangulation in solid lines.
}

\section{Geometric Relations for Unstructured Meshes}

For unstructured meshes, we need to determine the relation between a
point and a simplex (triangle or tetrahedron), and also some geometric
quantities, like the volume of a simplex and the surface area of its faces.
In this appendix, we give the formulas to calculated these
quantities generally in $n$-dimensional space.

Let $p_1, p_2, \cdots, p_{n+1}$ be $n+1$ distinct points in $n$-dimensional
space. The $n$-dimensional volume of the simplex $T$ with vertices
$p_1, \cdots, p_{n+1}$ is given by
\begin{equation}
\text{Vol} (p_1, \cdots, p_{n+1}) = \frac{1}{n!} \left | \text{Det}
\left ( \begin{array}{cccc}
1 & 1 & \cdots & 1 \\
p_1 & p_2 & \cdots & p_{n+1}
\end{array} \right ) \right | .
\end{equation}
Let $q$ be an arbitrary point in $n$-space.
If the simplex $T$ with vertices $p_1, \cdots, p_{n+1}$ is nondegenerate, i.e.,
if $\text{Vol} (p_1, \cdots, p_{n+1}) \ne 0$, the numbers, 
$b_1, b_2, \cdots, b_{n+1}$, satisfying
\begin{equation}
\left (
\begin{array}{cccc}
1 & 1 & \cdots & 1 \\
p_1 & p_2 & \cdots & p_{n+1}
\end{array}
\right )
\cdot
\left ( \begin{array}{c} b_1 \\ b_2 \\ \vdots \\ b_{n+1} \end{array} \right )
=
\left ( \begin{array}{c} 1 \\ q \end{array} \right ) 
\label{eqn:barycenter}
\end{equation}
are called {\it barycentric coordinates} of a point $q$ relative to simplex $T$.
It can be shown that,
\begin{equation}
b_k =
{\text{Det} (p_1, \cdots, q, \cdots p_{n+1}) \over
\text{Det} (p_1, \cdots, p_k, \cdots p_{n+1})} .
\end{equation}
Obviously, $b_k$ is a linear function of $q$.
Thus, $b_k$ indicates the position of $q$ relative to the hyperplane $H_k$
containing the facet of simplex $T$ opposite to vertex $p_k$.
$b_k=0$ when $q$ is in $H_k$, $b_k>0$ when $q$ is on the same side of $H_k$
from $p_k$, and $b_k<0$ when $q$ is on the opposite side of $H_k$ from $p_k$.
As a consequence, we know that
the point $q$ is inside simplex $T$ if and only if all $b_k \ge 0$.
The surface area vector $\vec{S}_k$ of hyperplane $H_k$ with
its direction pointing away from $p_k$ is
\begin{equation}
\vec{S}_k = - \text{Vol} (T) \nabla b_k(x) .
\end{equation}

It can be shown that the integration of function $f(x_1,\cdots,x_n)$ over the volume
of a simplex $T$ can be expressed as,
\begin{equation}
\int_T f(x,y,z) d^n x = \text{Vol}(T) n! 
\int_0^1 d b_1 \int_0^{1-b_1} d b_2 \cdots
\int_0^{1-b_1 \cdots - b_{n-1}} d b_n
f (x_1, \cdots, x_n) .
\end{equation}

If a simplex $T$ is non-degenerate, it has a unique circum-sphere $S$.
Given an arbitrary point $q$ in $n$-space, we can determine if $q$ is
inside, outside or on the sphere $S$ by the following function,
\begin{equation}
\text{InSphere} (q,T) = {
\text{Det}
\left ( \begin{array}{cccc}
1 & 1& \cdots & 1 \\
w_q & w_1 & \cdots & w_{n+1} \\
q & p_1 & \cdots & p_{n+1}
\end{array} \right )
\over
\text{Det}
\left ( \begin{array}{cccc}
1 & 1 & \cdots & 1 \\
p_1 & p_2 & \cdots & p_{n+1}
\end{array} \right )
} ,
\end{equation}
where $w_p = \sum_{i=0}^n x_{p,i}^2$.
$\text{InSphere} (q,T)>0$ when $q$ is inside $S$,
$\text{InSphere} (q,T)=0$ when $q$ is on $S$ and
$\text{InSphere} (q,T)<0$ when $q$ is outside $S$.
When calculating the value of $\text{InSphere}(q,T)$,
we should be aware of round-off errors (\citeNP{Barth95}),
because the result of triangulation could be wrong
due to floating point inaccuracies.
To avoid the problems caused by floating point round-off errors,
we calculate the above $\text{InSphere} (T,q)$ function using the
following formula instead,
\begin{equation}
\text{InSphere} (T,q) =
{
\text{Det}
\left ( \begin{array}{cccc}
w'_1 & w'_2 & \cdots & w'_{n+1} \\
p'_1 & p'_2 & \cdots & p'_{n+1}
\end{array} \right )
\over
\text{Det}
\left ( \begin{array}{cccc}
1 & 1 & \cdots & 1 \\
p'_1 & p'_2 & \cdots & p'_{n+1}
\end{array} \right )
{\displaystyle \min_{k=1,2,\cdots, n+1} ( w'_k ) }
} ,
\end{equation}
where $p'_k=p_k-q$ and $w'_k = \sum_{i=0}^n x^{'2}_{p,i}$.
The value of the above function is compared with a small number $\epsilon$,
instead of $0$, to determine the result of the sphere test.
For single precision floating operations, we use $\epsilon=10^{-4}$.
Our numerical experiments show that the above estimate is sufficient to
give the correct Delaunay triangulation.
Another way to avoid the round-off error is use exact
redundant expression calculation (see \citeNP{Barth95} for more discussions).
But there are a lot of extra calculations related to the exact
redundant expression \cite{FvW93}.

\section{Coefficient calculation in BGK formalism}

In this appendix, we will give detailed formulas to calculate the
coefficients $A^L_\beta$, $A^R_\beta$, $A^{GL}_\beta$, $A^{GR}_\beta$,
$B_\beta$, and other quantities for the BGK scheme described earlier
in the paper.

At the beginning of each time step, we know the fluid state at the two
ends of each edge $U^L$ and $U^R$. The interpolated fluid state
in the middle of the edge, $\bar{U}^L$ and $\bar{U}^R$, which are
interpolated from left side and right side respectively, can be
constructed from the SLIP (Symmetric LImited Positive) formulation
(\citeNP{Jameson95}), 
which is derived from the local extremum diminishing (LED) principle.
The constructed fluid state can be expressed as,
\begin{equation}
\bar U^L_\alpha = U^L_\alpha + \frac{1}{2} e^L_\alpha, \ \text{and} \ 
\bar U^R_\alpha = U^R_\alpha + \frac{1}{2} e^R_\alpha,
\end{equation}
where
$e_{\alpha, j} = L(\Delta U_{\alpha,j+1/2}, \Delta U_{\alpha,j-1/2})$
is the limited average,
$L(u,v)$ is a limiter,
and $\Delta U_{\alpha, j+1/2} = U_{\alpha,j+1} - U_{\alpha, j}$.
An example is the van Leer limiter,
$L(u,v) = \frac{2 u v}{u + v}$ when  $u \cdot v > 0$, and 
$L(u,v) = 0$ otherwise.
The equilibrium distribution functions $g_0^L$ and $g_0^R$ are
constructed from $\bar{U}_L$ and $\bar{U}^R$ respectively.

The macroscopic quantities are moments of distributions. We have
$\tilde U^L_\alpha(x) = \int \psi_\alpha g^L (1 + A^L_\beta \psi_\beta
x) d \vec{u} d \xi$.
For compatibility, we require $\tilde U^L_\alpha(x=-\frac{1}{2}) = U^L_\alpha$.
After some algebra, we get the solutions of the the coefficients
$A_\beta^L$ and $A_\beta^R$,
\begin{equation}
<\psi_\alpha \psi_\beta>^L A^L_\beta = e^L_\alpha \ \text{and} \ 
<\psi_\alpha \psi_\beta>^R A^R_\beta = e^R_\alpha .
\end{equation}
The notation $<\cdots>$ will be defined later in Appendix C.

The constructed fluid state at the middle of an edge $\bar{U}^G$ is
defined to be,
\begin{equation}
\bar{U}_\alpha^G \equiv \int \psi_\alpha f_0 d \vec{u} d \xi
= \int_{u_x>0} \psi_\alpha g_0^L d \vec{u} d \xi
+ \int_{u_x<0} \psi_\alpha g_0^R d \vec{u} d \xi .
\end{equation}
The equilibrium distribution $g_0^G$ is constructed from $\bar{U}^G$.
Taking the limit $t \rightarrow 0$,
we have
$\tilde U^G_\alpha(x) = \int \psi_\alpha g^G (1 + A^G_\beta \psi_\beta
x) d \vec{u} d \xi $.
For $x=\pm\frac12$, we require
$\tilde{U}_\alpha^G(-\frac12) = U_\alpha^L$ and
$\tilde{U}_\alpha^G(+\frac12) = U_\alpha^R$.
This gives the solutions of coefficients $A_\beta^{GL}$ and
$A_\beta^{GR}$.

The distribution functions $f(t,x,u)$ and $g(t,x,u)$ must be compatible
with each other. Conservation laws give the following compatibility
condition,
\begin{equation}
\int \psi_\alpha (f-g) d \vec{u} d \xi = 0 .
\end{equation}
Applying the integrated solution of $f(t,\vec{x},\vec{u})$
(equation \ref{eqn:BGK3}) to the above compatibility equation
and integrate over the whole time step $T$, we have,
\begin{eqnarray}
0 &=& \int \psi_\alpha d\Xi dt {1 \over \tau} \int_0^T (f-g) \nonumber \\
&=&
	- \left [ 1-e^{-T/\tau} \right ] U^G_\alpha
\nonumber \\ & &
	+ \left [ 2 \tau (1-e^{-T/\tau}) - T (1+e^{-T/\tau}) \right ]
	\left ( <u_x \psi_\alpha \psi_\beta>^G_{u>0} A^{GL}_\beta
	      + <u_x \psi_\alpha \psi_\beta>^G_{u<0} A^{GR}_\beta \right )
\nonumber \\ & &
	+ \left [-T + \tau (1-e^{-T/\tau}) \right ]
	<\psi_\alpha \psi_\beta>^G B_\beta
\nonumber \\ & &
	+ \left [ 1-e^{-T/\tau} \right ]
	\left ( <\psi_\alpha>_{u>0}^L
	      + <\psi_\alpha>_{u<0}^R \right )
\nonumber \\ & &
	- \left [ \tau-(T+\tau) e^{-T/\tau} \right ]
	\left ( A^L_\beta < u \psi_\alpha \psi_\beta >_{u>0}^L 
	      + A^R_\beta < u \psi_\alpha \psi_\beta >_{u<0}^R \right)
\label{eqn:Beqn}
\end{eqnarray}
These equations give the solutions to the coefficients $B_\beta$.
\comment{
where,
\begin{eqnarray}
k_1 &=& 1-e^{-T/\tau} \nonumber \\
k_2 &=& 2 \tau (1-e^{-T/\tau}) - T (1+e^{-T/\tau}) \nonumber \\
k_3 &=& -T + \tau (1-e^{-T/\tau}) \nonumber \\
k_4 &=& 1-e^{-T/\tau} \nonumber \\
k_5 &=& \tau-(T+\tau) e^{-T/\tau} \nonumber
\end{eqnarray}
}

For convenience, we give the formula to calculate
the time integrated flux functions,
\begin{eqnarray}
\lefteqn{ \int_0^T \vec{F}_\alpha(0,t) dt \equiv \int_0^T dt \int
\vec{u} \psi_\alpha f(t,0,\vec{u}) d \vec{u} d \xi } \nonumber \\
&=& \begin{array}[t]{l}
	\left [ T-\tau (1-e^{-T/\tau}) \right ] <u \psi_\alpha>^G \\
	+ \tau \left [ 2\tau (1-e^{-T/\tau}) - T (1+e^{-t/\tau}) \right ]
	  \left [ A^{GL}_\beta <u^2 \psi_\alpha \psi_\beta>^G_{u>0} 
		+ A^{GR}_\beta <u^2 \psi_\alpha \psi_\beta>^G_{u<0} \right ] \\
	+ \left [ T^2/2 - T \tau + \tau^2 (1-e^{-T/\tau}) \right ]
		  B_\beta <u \psi_\alpha \psi_\beta>^G \\
	+ \tau \left ( 1-e^{-T/\tau} \right )
	  \left [ <u \psi_\alpha>_{u>0}^L + <u \psi_\alpha>_{u<0}^R \right ] \\
	- \tau \left ( \tau-(T+\tau)e^{-T/\tau} \right )
	  \left [ A^L_\beta<u^2 \psi_\alpha \psi_\beta>_{u>0}^L 
		  + A^R_\beta <u^2 \psi_\alpha \psi_\beta>_{u<0}^R \right ]
	\end{array} .
\end{eqnarray}

The collision time $\tau$ can be derived from classical statistical
mechanics to be the mean free path divided by the rms velocity of atoms.
We use the following formula to estimate $\tau$
\begin{equation}
\tau = C_1 { \sqrt{\lambda} \over \rho }
   + C_2 \Delta T { |p^L - p^R| \over p^L + p^R },
\end{equation}
where $\Delta T$ is the time step and $C_1$, $C_2$ are constants. We take
$C_1=0.01$ and $C_2=1$ in our calculations.
The results are not sensitive to the choice of the actual values of $C_1$,
$C_2$.

\section{Velocity moments}

We define the moment of a quantity $w$ of the equilibrium state $g_0$ as
the following,
\begin{equation}
<w> \equiv \frac{1}{\rho} \int w g_0 d \vec{u} d \xi
= \frac{1}{\rho} \int w D e^{-\lambda [ (u-V)^2 + \xi^2 ]} d\vec{u} d\xi ,
\end{equation}
where $V$ is the macroscopic velocity,
$\lambda \equiv \rho / 2 p$ is the gas temperature,
and $D=\rho (\lambda/\pi)^{(K+N)/2}$
is the normalization factor.
Here, $K$ is the degree of the internal variable $\xi$ and $N$ is the space
dimension \cite{kxu93}.
For a polytropic gas, classical statistical mechanics gives $\gamma=(n+2)/n$,
where $n$ is the total number of effective degrees of freedom of the
molecule: thus a monoatomic gas has $n=3$, $\gamma=5/3$, and a diatomic gas
with two rotational degrees of freedom 
has $n=5$, $\gamma=7/5$.
For a flow in $N$-dimensional space, we have
$K = n-N = \frac{2}{\gamma-1} - N$.

Following the above definition, we obtain the iterative relation,
\begin{equation}
<u^{n+2}> = V <u^{n+1}> + {n+1 \over 2\lambda} <u^n> ,
\label{eqn:un_mom}
\end{equation}
and the following specific values of moments,
\begin{eqnarray}
<u^0> &=& 1 \\
<u^1> &=& V \\
<\xi^2> &=& {K \over 2 \lambda} \\
<\xi^4> &=& {K(K+2) \over 4 \lambda^2} \\
<\xi^6> &=& {K(K+2)(K+4) \over 8 \lambda^3}
\end{eqnarray}

For velocity moments involving integration over half of the velocity
space, the above iterative relation (equation \ref{eqn:un_mom}) still
holds true, except for the following first few moments,
\begin{eqnarray}
<u_x^0>_{u_x>0} &=& \frac12 {\rm erfc} (-\sqrt\lambda V_x) \\
<u_x^1>_{u_x>0} &=& V_x \frac12 {\rm erfc} (-\sqrt\lambda V_x)
	+ {e^{-\lambda V_x^2} \over 2 \sqrt{\pi\lambda} } \\
<u_x^0>_{u_x<0} &=& \frac12 {\rm erfc} (\sqrt\lambda V_x) \\
<u_x^1>_{u_x<0} &=& V_x \frac12 {\rm erfc} (\sqrt\lambda V_x)
	- {e^{-\lambda V_x^2} \over 2 \sqrt{\pi\lambda} } 
\end{eqnarray}

The moments of $u^n \psi_\alpha$ and $u^n \psi_\alpha\psi_\beta$ can be derived
from the moments of $u^n$ and $\xi^n$. We explicitly write them out for
reference.
\begin{eqnarray}
\lefteqn{
<u^n \psi_\alpha>
\equiv 
\frac{1}{\rho} \int u^n \psi_\alpha D e^{-\lambda (u-V)^2 + \xi^2} d\Xi
}
\\
& = & \left ( \begin{array}{c}
  <u^n> \\ <u^{n+1}> \\ \frac12 ( <u^{n+2}> + <u^n> <\xi^2> )
  \end{array} \right )
\nonumber \\
\lefteqn{
<u^n \psi_\alpha \psi_\beta> \equiv 
\frac{1}{\rho}
\int u^n \psi_\alpha \psi_\beta D e^{-\lambda (u-V)^2 + \xi^2} d\Xi
}
\\
& = & \left (
	\begin{array}{ccc}
	<u^n> & <u^{n+1}> & \frac{1}{2} <u^{n+2} + u^n \xi^2> \\
	<u^{n+1}> & <u^{n+2}> & \frac{1}{2} <u^{n+3} + u^{n+1} \xi^2> \\
	\frac{1}{2} <u^{n+2} + u^n \xi^2> & \frac{1}{2} <u^{n+3} +u^{n+1} \xi^2> &
	\frac{1}{4} <u^{n+4} + 2 u^{n+2} \xi^2 + u^n \xi^4> 
	\end{array}
	\right )
\nonumber
\end{eqnarray}

For moments of $u^n \psi_\alpha$ and $u^n \psi_\alpha\psi_\beta$
integrating over half velocity space, the above expressions are still good
except that one must be aware that $<u^0>$ may not equal to $1$
in the above formulas.

\section{Cooling Rates and Reaction Coefficients}

Various cooling processes relevant to primordial baryonic matter
have been included in the code.
Primordial gas is assumed consist essentially entirely of hydrogen
and helium.
The relevant cooling and heating processes are: bremsstrahlung emission,
collisional excitation of $H_0$ and $He^+$,
collisional ionization of $H_0$, $He^0$ and $He^+$,
radiative recombination of $H^+$, $He^+$ and $He^{++}$, 
and
dielectronic recombination of $He^+$.
Ionization equilibrium is assumed to determine the fraction in each
species.
The reaction rates $k_i$ (in unit of sec$^{-1}$) and the related cooling
rates $\Lambda_i$ (in unit of erg cm$^{3}$ sec$^{-1}$) are
listed below (c.f. \citeNP{Black81}, \citeNP{Cen92}, \citeNP{KWH96} and \citeNP{AAZN96}), (Note that $T_n \equiv T/10^n$K.)

\begin{enumerate}
\item $H^0 + e \rightarrow H^+ + 2 e$ 
$$k_1 = 5.85 \times 10^{-11} T^{1/2} e^{-157809.1/T} (1+T_5^{1/2})^{-1}$$
$$\Lambda_1 = 1.27 \times 10^{-21} T^{1/2} e^{-157809.1/T} (1+T_5^{1/2})^{-1} n_e n_{H^0}$$
\item $H^+ + e \rightarrow H^0 + \gamma$ 
$$k_2 = 8.40 \times 10^{-11} T^{1/2} T_3^{-0.2} (1+T_6^{0.7})^{-1}$$
$$\Lambda_2 = 8.70 \times 10^{-27} T^{1/2} T_3^{-0.2} (1+T_6^{0.7})^{-1}
n_e n_{H^+}$$
\item $He^0 + e \rightarrow He^+ + e$ 
$$k_3 = 2.38 \times 10^{-11} T^{1/2} e^{-285335.4/T} (1+T_5^{1/2})^{-1}$$
$$\Lambda_3 = 9.38 \times 10^{-22} T^{1/2} e^{-285335.4/T} (1+T_5^{1/2})^{-1} n_e n_{He^0} $$
\item $He^+ + e \rightarrow He^0 + \gamma$ 
$$k_4 = 1.5 \times 10^{-10} T^{-6.353}$$
$$\Lambda_4 = 1.55 \times 10^{-26} T^{0.3647} n_e n_{He^+}$$
\item $He^+ + e \rightarrow He^{++} + 2 e$ 
$$k_5 = 5.68 \times 10^{-12} T^{1/2} e^{-631515.0/T} (1+T_5^{1/2})^{-1}$$
$$\Lambda_5 = 4.95 \times 10^{-22} T^{1/2} e^{-631515.0/T} (1+T_5^{1/2})^{-1}
n_e n_{He^+}$$
\item $He^{++} + e \rightarrow He^+ + \gamma$ 
$$k_6 = 3.36 \times 10^{-10} T^{-1/2} T_3^{-0.2} (1+T_6^{0.7})^{-1}$$
$$\Lambda_6 = 3.48 \times 10^{-26} T^{-1/2} T_3^{-0.2} (1+T_6^{0.7})^{-1}
n_e n_{He^{++}}$$
\item $He^+ + 2e \rightarrow He^0 + e$ 
$$k_7 = 1.9 \times 10^{-3} T^{-1.5} e^{-470000.0/T} (1+0.3e^{-94000.0/T})$$
$$\Lambda_7 = 1.24 \times 10^{-13} T^{-1.5} e^{-470000.0/T} (1+0.3e^{-94000.0/T}) n_e n_{He^+}$$
\item collisional excitation of $H^0$
$$\Lambda_{8} = 7.50\times 10^{-19} e^{-118348.0/T} (1+T_5^{1/2}) n_e n_{H^0}$$
\item collisional excitation of $He^+$
$$\Lambda_{9} = 5.54\times 10^{-17} T^{-3.97} e^{-473638.0/T}
(1+T_5^{1/2})^{-1} n_e n_{He^+}$$
\item bremsstrahlung 
$$\Lambda_{ff} = 1.42 \times 10^{-27} g_{ff} T^{1/2} n_e ( n_{H^+} + n_{He^+} + 4 n_{He^{++}} ) $$
$$g_{ff} = 1.1 + 0.34 \exp[ -(5.5-\log T)^2/3.0 ]$$
\end{enumerate}

When an ultraviolet (UV) background radiation field is present, photoionization
of $H^0$, $He^0$ and $He^+$ is also included.
The photoionization rates are defined by
\begin{equation}
\Gamma_{\gamma i} \equiv \int_{\nu_i}^\infty \frac{4\pi J(\nu)}{h\nu}
\sigma_i(\nu) d \nu,
\end{equation}
where $J(\nu)$ is the intensity of the UV background, $\nu_i$ is the
threshold frequency and $\sigma_i(\nu)$ is the photoionization cross section
for species $i$.
The heating rate associated with photoionization is
\begin{equation}
{\mathcal{H}} = n_{H^0} \epsilon_{H^0} + n_{He^0} \epsilon_{He^0}
+ n_{He^+} \epsilon_{He^+} 
\ {\rm erg} {\rm cm}^{-3} {\rm sec}^{-1},
\end{equation}
where
\begin{equation}
\epsilon_i \equiv \int_{\nu_i}^\infty \frac{4 \pi J(\nu)}{h \nu}
\sigma_i(\nu) (h\nu - h \nu_i) dv.
\end{equation}

Besides these radiative cooling processes, we include inverse Compton
cooling off the microwave background. The inverse Compton cooling rate is
given by (\citeNP{IO86}),
\begin{equation}
\Lambda_C = 5.41 \times 10^{-36} n_e T (1+z)^4
{\rm erg} {\rm cm}^{-3} {\rm sec}^{-1} .
\end{equation}

\bibliographystyle{apj}
\bibliography{mnrasmnemonic,delaunay,lyman,reionization,books}

\end{document}